\documentclass[a4paper,12pt,english]{article}

\usepackage{amsfonts,bm,amssymb,euscript,array,babel,cite}
\usepackage[all,color]{xy}
\usepackage{relsize}
\usepackage{tikz}
\usepackage{url}
\usepackage{amsmath,amsthm}
\usepackage{mathtools}
\usepackage[]{graphicx}
\usepackage[T1]{fontenc}
\usepackage{framed}

\usepackage{empheq}
\usepackage{multirow}
\usepackage{hyperref}

\definecolor{darkgreen}{rgb}{0.0, 0.5, 0.0}

\usepackage{color}

\makeatletter
\DeclareRobustCommand\widecheck[1]{{\mathpalette\@widecheck{#1}}}
\def\@widecheck#1#2{%
	\setbox\z@\hbox{\m@th$#1#2$}%
	\setbox\tw@\hbox{\m@th$#1%
		\widehat{%
			\vrule\@width\z@\@height\ht\z@
			\vrule\@height\z@\@width\wd\z@}$}%
	\dp\tw@-\ht\z@
	\@tempdima\ht\z@ \advance\@tempdima2\ht\tw@ \divide\@tempdima\thr@@
	\setbox\tw@\hbox{%
		\raise\@tempdima\hbox{\scalebox{1}[-1]{\lower\@tempdima\box
				\tw@}}}%
	{\ooalign{\box\tw@ \cr \box\z@}}}
\makeatother

\newcommand\ringring[1]{%
	{
		\mathop{\kern0pt #1}\limits^{
			\vbox to-1.85ex{
				\kern-2ex 
				\hbox to -1pt{\hss\normalfont\kern.1em \r{}\kern-.45em \r{}\hss}%
				\vss 
			}
		}
	}
}

\newcommand\ringringring[1]{%
	{
		\mathop{\kern0pt #1}\limits^{
			\vbox to-1.85ex{
				\kern-2ex 
				\hbox to -1pt{\hss\normalfont\kern.1em \r{}\kern-.45em\r{}\kern-.45em \r{}\hss}%
				\vss 
			}
		}
	}
}

\makeatletter
\newcommand{\doublewidetilde}[1]{{%
  \mathpalette\double@widetilde{#1}%
}}
\newcommand{\double@widetilde}[2]{%
  \sbox\z@{$\m@th#1\widetilde{#2}$}%
  \ht\z@=.9\ht\z@
  \widetilde{\box\z@}%
}
\makeatother

\newsavebox{\mysaveboxM}
\newsavebox{\mysaveboxT}

\makeatletter

\newcommand{\dd}{\mathrm{d}}

\newcommand{\w}{\wedge}

\newcommand{\be}{\begin{equation}}
\newcommand{\ee}{\end{equation}}

\def\nn{\nonumber}
\def \bea{\begin{eqnarray}} 
\def\eea{\end{eqnarray}}
\def\bse{\begin{subequations}}	
\def\ese{\end{subequations}}
\def\bal{\begin{align}} 
\def\eal{\end{align}}
\newcommand{\mf}{\mathfrak}
\def\mc{\mathcal}

\def\bi{\begin{itemize}} 
\def\ei{\end{itemize}}

\makeatother

\def\a{\alpha}  \def\g{\gamma}   
\def\e{\epsilon} \def\vare{\varepsilon}
  \def\h{\eta} \def\k{\kappa}
\def\l{\lambda}  \def\m{\mu}
\def\n{\nu} \def\o{\omega}   
\def\s{\sigma} \def\S{\Sigma}  \def\th{\theta}

\def\R{{\mathbb R}}  \def\N{{\mathbb N}}  
 
 \def\Z{{\mathbb Z}} 

\def\one{\mbox{1 \kern-.59em {\rm l}}}

\numberwithin{equation}{section}

\begin{document}

\makeatother
\parindent=0cm
\renewcommand{\title}[1]{\vspace{10mm}\noindent{\Large{\bf #1}}\vspace{8mm}} \newcommand{\authors}[1]{\noindent{\large #1}\vspace{5mm}} \newcommand{\address}[1]{{\itshape #1\vspace{2mm}}}

\begin{titlepage}

\begin{flushright}
 RBI-ThPhys-2020-52
\end{flushright}

\begin{center}

\title{ {\large {Duality and higher Buscher rules \\[4pt] in $p$-form gauge theory and linearized gravity}}}

  \authors{\large Athanasios {Chatzistavrakidis}, Georgios Karagiannis, Arash Ranjbar}
 
 
  \address{  Division of Theoretical Physics, Rudjer Bo\v skovi\'c Institute \\ Bijeni\v cka 54, 10000 Zagreb, Croatia 
 }

\vskip 1cm

\vskip 1cm

\begin{abstract}

We perform an in-depth analysis of the transformation rules under duality for couplings of theories containing multiple scalars, $p$-form gauge fields, linearized gravitons or $(p,1)$ mixed symmetry tensors. Following a similar reasoning to the derivation of the Buscher rules for string background fields under T-duality, we show that the couplings for all classes of aforementioned multi-field theories transform according to one of two sets of duality rules. These sets comprise the ordinary Buscher rules and their higher counterpart; this is a generic feature of multi-field theories in spacetime dimensions where the field strength and its dual are of the same degree. Our analysis takes into account topological theta terms and generalized $B$-fields, whose behavior under duality is carefully tracked. For a 1-form or a graviton in 4D, this reduces to the inversion of the complexified coupling or generalized metric under electric/magnetic duality. Moreover, we write down an action for linearized gravity in the presence of $\theta$-term from which we obtain previously suggested on-shell duality and double duality relations. This also provides an explanation for the origin of 
theta in the gravitational duality relations as a specific additional sector of the linearized gravity action.  

\end{abstract}

\end{center}

\vskip 2cm

\end{titlepage}

\tableofcontents

\setcounter{footnote}{0}

\section{Introduction} 

Target space and strong/weak coupling dualities are prominent examples of the equivalence between two seemingly different pictures of physical theories. In their most elementary manifestation, a single compact scalar in 2 dimensions exhibits an $R\leftrightarrow 1/R$ T-duality, where the radius of the circular direction in which the scalar propagates is inverted, whereas a single Abelian 1-form in 4 dimensions exhibits an $e\leftrightarrow 1/e$ electric/magnetic S-duality, where the dimensionless electromagnetic coupling is inverted. Although conceptually rather different, these two simple dualities share the above ``inversion of coupling'' property. In both cases, this is a remnant of Hodge duality of the corresponding field strengths in self-dual dimensions, or in other words, of the respective conserved currents of their (generalized) global symmetries.{\footnote{Recall that classical electrodynamics has a $U(1)_e\times U(1)_m$ ``1-form'' global symmetry with corresponding currents the Faraday tensor and its dual \cite{Gaiotto:2014kfa}.}}  

T-duality becomes even more pronounced and interesting for multiple scalar fields $X^{M}, M=1,\dots,d,$ in 2 dimensions. In that case, one considers a nonlinear sigma model on a two-dimensional spacetime $\S$ with a $d$-dimensional target spacetime $M$ where the scalar fields play the role of components of a map from $\Sigma$ to $M$. Such a two-dimensional field theory is nonlinear since the corresponding couplings are functions of the scalar fields, thereby acquiring a geometric interpretation as a Riemannian metric on $M$. As is well known, although for a single scalar there is no topological $\theta$ term, multiple scalars have topological couplings $B_{MN}(X)$, corresponding to the Kalb-Ramond 2-form, which complement the ``Riemannian metric'' couplings $G_{MN}(X)$. Although it is not customary to call the $B$-field term a $\theta$-term, it will prove useful to adopt this unconventional terminology in this paper. In this more general setting, T-duality mixes $G_{MN}$ and $B_{MN}$ and leads to the celebrated Buscher rules, which generalize the $R\leftrightarrow 1/R$ picture, thus providing two distinct spacetime backgrounds which are physically equivalent. 

The Buscher rules are typically derived using either a procedure where one or multiple isometries are gauged, or by considering a type of first order parent action with Lagrange multipliers and algebraically eliminating different fields via their equations of motion. In the present paper we employ the second viewpoint. On this basis, one can ask the question:
\begin{itemize}
	\item \it{How stringy are the Buscher rules?}
	\end{itemize} 
Let us explain what we mean. Given the aforementioned similarity between the behavior of couplings under T-duality for a single scalar in 2D and electric/magnetic duality for a single Abelian 1-form in 4D, one could wonder whether this similarity persists for multiple fields and, for that matter, multiple $p$-forms in $2p+2$ dimensions, where the corresponding field strength and its Hodge dual are of the same degree.\footnote{As already mentioned, the dualities for scalars and 1-forms are conceptually different; this is generalized to even ($p=2k$) vs. odd ($p=2k-1$) differential forms in respectively $4k+2$ and $4k$ dimensions \cite{Julia:1980gr,Julia:1997cy,Deser:1997mz}. Here we only focus on the behavior of the background fields.} To address this, we are led to consider theories describing the dynamics of $d$ Abelian $p$-forms $A^M=1/p!\, A^M_{\mu_1...\mu_{p}}\,\dd x^{\mu_1}\w...\w \dd x^{\mu_{p}}, M=1,\dots, d,$ through the action:
\bea \label{pform}
S[A^{M}]=-\frac 12 \int_{\S_{2p+2}} \left(G_{MN}\, \dd A^{M}\w\ast\, \dd A^{N}+B_{MN}\, \dd A^{M}\w\dd A^{N}\right)\,.
\eea 
 One could view such theories, for example Maxwell theory in 4D or 2-form gauge theory in 6D and their multi-field generalizations, as sigma models from $(2p+2)$-D spacetime $\S_{2p+2}$ to a $d$-dimensional target space of fields ${\cal M}$, see for example the discussion in Ref. \cite{Gruetzmann:2014ica} for towers of differential forms. One can immediately note a few differences compared to the scalar case. First, the couplings could be field-dependent as long as some physical requirements hold, such as gauge invariance. Indeed, the above action may be considered as a sector of a theory containing additional propagating fields, typically scalars on which the couplings may depend, as is the case in supergravity theories. Second, the topological term contains both diagonal and off-diagonal contributions, a.k.a. ordinary $\theta$-terms and $B$-fields respectively. For even $p$ only the latter exist. Third, $B_{MN}$ is a 2-form in target space for even $p$, but instead it is symmetric for odd $p$. Finally, unlike the 2D case, where the string miracle leads to the Einstein equations in target space after quantization, we do not anticipate a similar interpretation in the general case. 
 
  Nevertheless, such theories may be dualized within the general procedure mentioned above, i.e. starting from a parent action and integrating out fields. The result of this procedure is two classes of higher Buscher rules, one being the standard Buscher rules for 2D scalars, now extended to twisted self-dual even $p$-forms, and one being the standard S-duality rules for 4D 1-forms, now extended to multiple fields and all twisted self-dual odd $p$-forms. We employ the terminology ``twisted self-duality'' as in \cite{Cremmer:1998px,Bunster:2011qp}, to indicate the difference between strictly self-dual forms, for instance $F=\ast\, F$ for a 2-form $F$ in 4D, and distinct dual forms, as in $\widehat{F}=\ast\, F$, with $\widehat{F}$ being related to a separate, magnetic 1-form potential.
  As such, Buscher rules turn out to not be stringy at all, but instead they are a general consequence of duality for differential forms in self-dual dimensions.
  
  Challenging the ``differential form'' part of the latter statement, one could wonder what happens in theories with a single or multiple graviton(s), at least at the linearized level. To address this, one may recall that linearized general relativity, essentially yielding the massless Fierz-Pauli action, exhibits a gravitational analogue of duality as a Hodge duality of linearized Riemann tensors, leading to the concept of dual graviton. This can be demonstrated both on-shell \cite{Hull:2000zn,Hull:2001iu}, namely at the level of equations of motion, and off-shell \cite{West:2001as,Boulanger:2003vs,Henneaux:2004jw}, namely at the level of the Lagrangian. Note that both approaches can be elegantly formulated in the framework of graded geometry, which is extremely useful when dealing with mixed symmetry tensor fields like the graviton \cite{Chatzistavrakidis:2019len}. We will employ this unifying formalism for this part of our analysis in the main text, briefly explaining its relation to the component approach too. 
  
  Notably, there exists a further generalization of the above concept of duality in gravity, more akin to our previous discussion, where a gravitational $\theta$-term plays a role. This was suggested in \cite{Hull:2001iu} and it is based on the on-shell duality relations 
  \bea \label{HullDR}
  \widehat{R}&=&\frac 1{g^2}\ast R -\theta R\,, \\ 
  \widecheck{R}&=&-\frac{2\theta}{g^2}\ast R+\left(\theta^2-\frac{1}{g^4}\right)R\,, \label{HullDR2}
  \eea
where $R$, $\widehat{R}$ and ${\widecheck R}$ are dual linearized Riemann tensors. The first of these duality relations corresponds to a Hodge duality of only one of the two slots of the graviton, thought of as a degree-$(1,1)$ tensor field, whereas the second one corresponds to Hodge duality along both of its slots, usually termed ``double duality''. However, to our knowledge, there has been no off-shell action from which these duality relations can be obtained once we go on-shell. Therefore, the second question we would like to address is:
\begin{itemize} 
	\item \it{Which parent action implements the duality relations \eqref{HullDR} and \eqref{HullDR2} and which are its dual children theories?}
	\end{itemize}
Note that at the on-shell level, it is not clear what $\theta$ corresponds to. This is clarified by the off-shell approach that we follow. We find that the parent action that implements these duality relations leads to two dual theories that apart from the kinetic (Fierz-Pauli) term contain a specific topological $\theta$-term. One might expect that this role is played by the linearized Pontryagin density, with the gravitational Chern-Simons form as its boundary. However, this would result in more derivatives than one requires for the on-shell duality relation. Indeed, we propose that the relevant topological $\theta$-term is what was recently suggested in \cite{Chatzistavrakidis:2020wum} in the context of modified gravitoelectromagnetism and its origin may be traced in the Nieh-Yan invariants of gravity with torsion \cite{Nieh:1981ww,Chandia:1997hu,Li:1999ue}. This is the second main result of the present paper. Note also that in the absence of $\th$, the duality relation \eqref{HullDR2} becomes rather trivial, since the two would-be dual gravitons are related algebraically rather than via a non-local relation. This was observed and studied in \cite{Henneaux:2019zod}. However, a side result of our analysis is that in the presence of such a theta term, double duality does not lead to an algebraic relation between the two dual fields.

Finally, one may even ask:
\begin{itemize}
	\item \it{What is the gravitational analogue of the Buscher rules?  }
	\end{itemize}
In other words, considering a theory of multiple gravitons,\footnote{Multi-graviton theories were considered for example in  \cite{Boulanger:2000rq}, where the question of constructing interacting theories was addressed. Here, we restrict our analysis to the free field case.} of the same form as \eqref{pform}, we determine a first order parent action that implements the multiple field generalization of the duality relations \eqref{HullDR} and \eqref{HullDR2}. We argue that upon eliminating different fields in the parent action, one can identify the two dual theories whose couplings are related by the same higher Buscher rules as for the 1-form. We also discuss how this generalizes to the case of higher mixed symmetry tensor fields of degree $(p,1)$.

The paper is organized as follows. In Section \ref{sec2}, we set the stage for our analysis by recalling some well-known materials on the duality for scalars in 2D and the derivation of the Buscher rules for background fields. In Section \ref{sec3}, this is generalized to Abelian 1-forms with topological $\theta$-terms, both in the single and multiple field(s) cases. This leads to the implementation of the corresponding dualities in an alternative way to the Hamiltonian approach of \cite{Bunster:2011aw}. In particular, it leads to a set of Buscher-like rules for the couplings, containing the allowed generalized $\theta$-terms and $B$-fields. In Section \ref{sec4}, we discuss in detail how this duality procedure at the level of the action can be used for a single graviton when $\theta$-terms are included, essentially implementing Hull's  duality relations \eqref{HullDR} and \eqref{HullDR2} for the dual and double dual graviton off-shell. In this analysis, we employ a graded-geometric formalism, which is introduced in Section \ref{sec41}. We also discuss the meaning of the $\theta$-term in this gravitational context. Section \ref{sec5} contains the generalization of the previous statements for multiple $p$-forms, multiple gravitons and higher mixed symmetry tensor fields of degree $(p,1)$. We show that essentially there is a single, universal parent action for all cases and the duality transformations for the couplings fall in one of two sets of Buscher rules, identified with the ones for scalars and 1-forms respectively. We also discuss in a unified manner the orthogonal or symplectic groups that underline the case of performing multiple dualities in Section \ref{sec52}. Section \ref{sec6} contains our conclusions. Some details of computations used in Section \ref{sec4} are provided in Appendix \ref{appa}. 
	
\section{Duality for scalars in 2D}
\label{sec2}

In this section, we briefly review the basic elements of T-duality for background fields in theories of periodic scalars in 2D, essentially recalling the derivation of the Buscher rules. Although this material is well known, this discussion will set the stage for our approach in the following sections, where we discuss higher $p$-form fields and gravitons.  

\subsection{Single scalar and $R \leftrightarrow 1/R$ duality} 
\label{sec21}

Consider a single periodic scalar field $X$ with period $2\pi$. It is well known that in two dimensions this is the setting for T-duality, which amounts to radius inversion of the corresponding circular directions in which the original and dual scalar fields propagate. Indeed, following the discussion in \cite{Hori:2000kt}, we consider the sigma model from a Lorentzian worldsheet $\S_2$ to a circle $S^{1}$ of radius $R$,{\footnote{Note that a Euclidean version is considered in \cite{Hori:2000kt}, which leads to some differences in the formulas. The action functionals depend on $\gamma_{\m\n}$ too, but we do not denote this dependence explicitly. }}
\be \label{SX}
S[X]=-\frac 1{2}\int_{\S_2} R^2\sqrt{\g}\,\g^{\mu\nu}\,\partial_{\mu}X\,\partial_{\nu}X\,\dd^2\sigma=-\frac 1{2}\int_{\S_2} R^{2}\,\dd X\w\ast \,\dd X\,,
\ee 
where $\s^{\mu},\mu=1,2$ are local coordinates on $\S_2$, $\g_{\mu\nu}$ are the components of the worldsheet metric, $\g$ its determinant and we set the string slope parameter $2\pi\a'=1$. Here and in the following, we mostly use differential form notation and the metric $\g_{\mu\nu}$ will not appear explicitly in the various formulas. Note that both $\dd X=\partial_{\mu}X\dd\s^{\mu}$, the ``field strength'' 1-form of the scalar field $X$, and its Hodge dual $\ast\dd X$ are 1-forms in 2D, where $\ast$ is the two-dimensional Hodge star operator that squares to $+1$ in the 2D Lorentzian case. 

In order to uncover the duality, we can consider a first order action which is a functional of $X$ and an additional, independent 1-form $F$. Alternatively, instead of $X$, one can start with another periodic scalar, say $\widehat{X}$, which will eventually correspond to the dual field. The two ways of approaching the problem are equivalent and here we employ the second (note that in \cite{Hori:2000kt}, the authors use $X$ instead.) Thus the parent action we consider is 
\be \label{parent scalar}
{\cal S}[F,\widehat{X}]=-\frac 1{2}\int_{\S_2}\, R^2 \, F\w\ast \,F+\int_{\S_2} \, F\w\dd\widehat{X}\,. 
\ee 
Integrating out $\widehat{X}$ and taking care of the global properties of the model (see \cite{Hori:2000kt} for the complete treatment) the original second order action \eqref{SX} with target $S^1$ of radius $R$ is recovered. On the other hand, integrating out the 1-form $F$ via its field equation 
\be 
F=\frac 1{R^2}\ast\dd\widehat{X}\,,
\ee 
the resulting dual second order action is 
\be 
\widetilde{S}[\widehat{X}]=-\frac 1{2}\int_{\S_2}\, \frac 1{R^2}\,\dd\widehat{X}\w\ast\,\dd \widehat{X}\,,
\ee 
which corresponds to a model for a compact scalar on a circle of radius $1/R$. This is how T-duality is realised as an equivalence of sigma models for a single scalar, specifically
\be 
S[X] \, \xleftarrow{\,\widehat{X} \, \text{on-shell}} \, {\cal S}[F,\widehat{X}] \, \xrightarrow{\,F \, \text{on-shell}} \, \widetilde{S}[\widehat{X}]\,.
\ee 
 Essentially it corresponds to an exchange of its Bianchi identity ($\dd^2 X=0$) and its equation of motion ($\dd\ast\dd X=0$) (cf. \cite{Duff:1990hn}).

\subsection{Multiple scalars and Buscher rules}
\label{sec22}

A trivial observation, which will be relevant in the broader context of the following sections, is that there is no topological term for a single scalar in two dimensions, simply due to degree reasons. One could say that there is no ``theta term''. This changes, however, when multiple scalar fields are considered. This is anyway the case in string models, where the scalars are 26 or 10 depending on whether we deal with bosonic strings or superstrings. Here we restrict to the bosonic string case, but we are going to be cavalier about the number of scalars by taking it to be some number $d$ that corresponds to the dimension of the target space. This will assume a critical value upon quantization as usual. The sigma model for multiple scalar fields $X^{M}$ is given by the action functional 
\be \label{SXmu}
S[X^{M}]=-\frac 1{2}\int_{\S_2} \left(G_{MN}(X)\,\dd X^{M}\w\ast\,\dd X^{N}+B_{MN}(X)\,\dd X^{M}\w\dd X^{N}+\cdots\right)\,,
\ee 
where the ellipses represent higher order terms, e.g. the dilaton coupling. Here scalar fields $X^{M}, M=1,\dots, d$, are thought of as the components of a map $X:\S_2\to  M_{d}$ from the worldsheet to a $d$-dimensional target space $M_{d}$. At the level of the sigma model per se, the symmetric and antisymmetric tensors $G_{MN}$ and $B_{MN}$ respectively, are the field-dependent couplings of the scalar field theory in two dimensions. Moreover, they directly generalize the coupling in the single scalar case ($M=1$), which was simply $G_{11}=R^{2}$, while the topological term was trivially absent.

T-duality for the background fields $G_{MN}$ and $B_{MN}$ can be investigated using the general sigma model \eqref{SXmu}, at least under certain conditions. The inversion of the radius of the single scalar case, or equivalently the statement that $G_{11}$ would go to $G^{11}$, where $G^{MN}$ is the inverse of $G_{MN}$, is proliferated to the famous Buscher rules \cite{Buscher:1987sk}. The procedure is presented in detail in many places and textbooks. We just mention in brief a few steps for reference in the following sections. First, assuming that the manifold $M_d$ has an isometry
generated by a Killing vector field $\rho$ such that 
\be 
{\cal L}_{\rho}G=0 \qquad \text{and} \qquad {\cal L}_{\rho}B=\dd\beta\,,
\ee   
where $\beta$ is an arbitrary 1-form, one may choose adapted coordinates $X^{M}=(X^{m},X), m=1,\cdots,d-1,$ such that the background fields are independent of $X\equiv X^{d}$. One may then rewrite the action \eqref{SXmu} in these adapted coordinates and eventually write down a first order action in terms of a scalar field $\widehat{X}$ and a 1-form $F$ as before, namely
\bea \label{S1}
{\cal S}[X^{m},\widehat{X},F]&=&-\frac 1{2}\int_{\S_2} \left(G_{mn}\,\dd X^{m}\w\ast\,\dd X^{n}+2\, G_{dm}\,F\w\ast\, \dd X^{m}+G_{dd}\, F\w\ast F\right) \nn\\[4pt] 
&&-\frac 1{2}\int_{\S_2} \left(B_{mn}\,\dd X^{m}\w\dd X^{n}+2B_{dm}\,F\w\dd X^{m}- 2 F\w\dd\widehat{X}\right)\,.
\eea   
Integrating out $\widehat{X}$ results in the Bianchi identity $\dd F=0$ and therefore locally $F=\dd X$.\footnote{At least after taking care of global issues that arise when the worldsheet has non-contractible cycles---see \cite{Plauschinn:2018wbo} for a detailed discussion.} In that case, \eqref{S1} reduces to \eqref{SXmu} in adapted coordinates. On the other hand, the field equation for the 1-form $F$ reads 
\be \label{DRmultiscalar}
F=\frac 1{G_{dd}}\left(\ast\,\dd\widehat{X}-G_{dm}\dd X^{m}-B_{dm}\ast\dd X^{m}\right)\,.
\ee 
Inserting this equation in the action \eqref{S1} and introducing the notation $\widehat{X}^{M}=(X^{m},\widehat{X})$ leads to a dual second order action of the desired form 
\be
\widetilde{S}[\widehat{X}^{M}]=-\frac 1{2}\int_{\S_2} \left(\widetilde{G}_{MN}(\widehat{X})\,\dd \widehat{X}^{M}\w\ast\,\dd \widehat{X}^{N}+\widetilde{B}_{MN}(\widehat{X})\,\dd \widehat{X}^{M}\w\dd \widehat{X}^{N}\right)\,,
\ee 
where the new couplings are given by the Buscher rules 
\bea 
&&\widetilde{G}_{dd}=\frac 1{G_{dd}}\,,\qquad \widetilde{G}_{md}=\frac {B_{md}}{G_{dd}}\,, \qquad \widetilde{B}_{md}=\frac{G_{md}}{G_{dd}}\,, \nn\\[4pt] 
&& \widetilde{G}_{mn}=G_{mn}-\frac {G_{md}G_{nd}-B_{md}B_{nd}}{G_{dd}}\,,\qquad \widetilde{B}_{mn}=B_{mn}-\frac {B_{md}G_{nd}-G_{md}B_{nd}}{G_{dd}}\,. \label{BuscherX}
\eea 
Furthermore, these rather complicated expressions for the transformation of couplings under T-duality acquire a more transparent form in terms of the generalized metric 
\be 
E_{MN}=G_{MN}+B_{MN}\,.
\ee 
What is more, this facilitates the generalization of the Buscher rules for multiple Killing directions. Indeed, in adapted coordinates $X^M=(X^m,X^i)$ with $X^i$ being the spatial Killing directions along which duality is performed, the Buscher rules may be written as \cite{Giveon:1994fu} 
\bea \label{buschergenmetric}
\widetilde{E}_{mn}=E_{mn}-E_{mi}E^{ij}E_{jn}\,,\quad \widetilde{E}_{mi}=E_{mj}E^{ji}\,, \quad \widetilde{E}_{ij}=E^{ij}\,,
\eea  
where $E^{ij}$ is the inverse of $E_{ij}$.
Although we do not provide the details for this multi-isometry generalization here, we will revisit these expressions in a more general context in Section \ref{sec52}.

Finally, note that two-dimensional scalars are special. While from the worldsheet perspective T-duality relates different string backgrounds that correspond to the same conformal field theory, the story is  exciting from the target space perspective too. This is mainly due to the richness of string theory and in particular due to the celebrated fact that the vanishing of the beta functions of the nonlinear sigma model corresponds to the Einstein equations for the low energy field theory in target space. In other words, the space where the couplings of the two-dimensional theory live is the target space and it is meaningful in itself as a physical spacetime. This is not expected to be the case in general, see however \cite{Bonezzi:2020jjq} for a derivation of Einstein equations from the quantization of the ${\cal N}=4$ superparticle. Nevertheless, one can still pose questions regarding duality of higher degree (gauge) fields in higher dimensions and the properties of the corresponding space of couplings. This is what we address in the following sections. 

\section{Duality for Abelian 1-Forms in 4D}
\label{sec3}

In this section, we perform a similar analysis to the previous one, this time for Abelian 1-forms in four dimensions. In this case, the relevant duality is the electric/magnetic S-duality of Maxwell electrodynamics, first suggested by O. Heaviside \cite{Heaviside}. Although the off-shell implementation of this duality is known, we are going to take it further in two directions. First, we include the corresponding theta term, which is related to the Abelian Chern-Simons form on the boundary, and second we generalize the analysis to multiple Abelian 1-forms with the goal of determining the analog of the Buscher rules in that case. 

\subsection{Single Maxwell field and  $\tau\leftrightarrow - 1/\tau$ duality}
\label{sec31}

On-shell electric/magnetic duality in Maxwell's electrodynamics amounts to the exchange of field equations and Bianchi identities. In differential form notation this becomes particularly transparent since the former are $\dd\ast F=0$, whereas the latter are $\dd F=0$, where $F=\dd A$ is now the 2-form field strength of the gauge potential $A$. Obviously, the exchange $F\leftrightarrow \ast F$ leaves the above system invariant. In fact, in the classical theory one can consider a continuous rotation in the plane of $F$ and $\ast F$, the resulting symmetry being $SO(2)$. Note that this symmetry can also be realised off shell \cite{Deser:1976iy,Deser:1981fr,Bunster:2011aw}.  

The inversion of coupling under electric/magnetic duality in Maxwell theory can be seen in the very same way as the inversion of radius for the single scalar in two dimensions. The underlying mathematics is identical, given that the Hodge dual of a 2-form in 4 dimensions is again a 2-form. Indeed, starting with the Maxwell action for the gauge field $A=A_{\m}\dd x^{\m}$, $x^{\m}$ being coordinates in 4D spacetime $\S_4$, 
\be \label{SA}
S[A]=-\frac 1{4e^2} \int \dd^4x F_{\m\n}F^{\m\n}=-\frac 1{2e^{2}} \int_{\S_4} \dd A\w\ast \,\dd A\,,
\ee 
we may introduce a first order action in terms of an independent{\footnote{Meaning that $F$ is not given as $\dd A$ at this stage, although we use the same symbol.}} 2-form $F$ and a 1-form $\widehat{A}$---which will become the magnetic gauge potential---that reads as
\be 
{\cal S}[F,\widehat{A}]=-\frac 1{2e^2}\int_{\S_4} F\w\ast F-\frac 1{4\pi}\int_{\S_4} F\w \dd \widehat{A}\,.
\ee  
Integrating out the dual 1-form $\widehat{A}$ from the first order action results in the original second order action \eqref{SA}, up to a total derivative term. On the other hand, integrating out $F$ via its field equation---note that $\ast^2 F=-F$ in four-dimensional Lorentzian space with mostly plus convention for the metric--- one finds 
\be 
F=\frac {e^2}{4\pi}\ast\dd\widehat{A}\,,
\ee 
which leads to the dual second order action 
\be 
\widetilde{S}[\widehat{A}]=-\frac 1{2\widetilde{e}^2}\int_{\S_4} \dd\widehat{A}\w\ast\,\dd\widehat{A}\,,
\ee 
where the dual coupling is related to the original one via inversion, specifically
\be \label{e1e}
\widetilde{e} = \frac {4\pi}{e}\,.
\ee 
Note that the same conclusion holds when the theory is coupled to background fields, in which case one may trace the 't Hooft anomaly of the electric and magnetic generalized (1-form) global symmetries of Maxwell theory through duality \cite{Cordova:2018cvg}.

Up to this point, we ignored the additional theta term that can be added in Maxwell theory in 4 dimensions. Although in general this term may be ignored, since it does not affect the quantization of the theory in the same way as the corresponding term in Yang-Mills theory, one should also take into account that such an Abelian theta term is important in physical contexts such as topological insulators. In particular, it plays an important role in duality of such physical systems \cite{Karch:2009sy,Mathai:2015raa}, and therefore it is meaningful to consider it in our analysis. The procedure which was described above works even in the presence of a theta term although necessary modifications are needed that we discuss here. Consider the inclusion of the topological term in the action of a single Maxwell field, 
\be 
S_{\theta}[A]=-\frac 1{2e^2}\int_{\S_4} \dd A\w\ast \,\dd A+\frac {\theta}{16\pi^2}\int_{\S_4} \dd A\w \dd A\,.
\ee 
Moreover, note that $\theta$ is a periodic variable and it is defined only modulo $2\pi$. Recall that in the physics of topological insulators, the relevant values of $\theta$ are $0$ and $\pi$, since time-reversal invariance is respected.

Next, we perform the dualization as before, but this time in the presence of the topological term. To this end, we write down the first order action, again in terms of the independent 2-form $F$ and the magnetic 1-form $\widehat{A}$, which is given as 
\bea \label{maxparent}
{\cal S}_{\theta}[F,\widehat{A}]=-\frac 1{2e^2}\int_{\S_4} F\w\ast F+\frac {\theta}{16\pi^2}\int_{\S_4} F\w F-\frac 1{4\pi}\int_{\S_4} F\w\dd \widehat{A}\,.
\eea 
As in all previous cases, the integration of the dual 1-form leads to the original first order action, whereas the integration of $F$ leads to the dual theory. Let us examine this dual theory in this case. The field equation for $F$ is 
\be 
\dd\widehat{A}=4\pi\left(-\frac 1{e^2}\ast F+\frac{\theta}{8\pi^2}F\right)\,. 
\ee 
We observe that it depends both on $F$ and its Hodge dual. Its solution for $F$ is
\be \label{FAtheta}
F=u\, \dd\widehat{A}+ v \ast\dd\widehat{A}\,.
\ee 
where
\be 
u=\frac{\theta e^4}{32\pi^3}\left(1+\left(\frac{\theta e^2}{8\pi^2}\right)\right)^{-1} \qquad \text{and} \qquad v=\frac{e^2}{4\pi}\left(1+\left(\frac{\theta e^2}{8\pi^2}\right)\right)^{-1}\,.
\ee 
It is then a straightforward calculation to insert $F$ from \eqref{FAtheta} in the first order action and determine the dual action, which has the general form 
\be 
\widetilde{S}_{\theta}[\widehat{A}]=-\frac 1{2\widetilde{e}^2}\int_{\S_4} \dd \widehat{A}\w\ast\, \dd \widehat{A}+\frac {\widetilde\theta}{16\pi^2}\int_{\S_4} \dd \widehat{A} \w \dd \widehat{A} \,.
\ee 
The new coupling and $\theta$ parameter are given in terms of the original ones as 
\bea 
\widetilde{e}^2=\frac{\theta^2e^4+64\pi^4}{4\pi^2e^2}\,,
\qquad
\widetilde{\theta}=-\frac {4\pi^2\theta e^4}{\theta^2e^4+64\pi^4}\,.
\eea 
This result generalizes the inversion \eqref{e1e}, which is obviously recovered when we set $\theta=0$. Moreover, one notices that this is the standard electric/magnetic duality even though it is not easy to see from these variables. Indeed, defining the usual complexified coupling 
\be 
\tau=\frac{\theta}{2\pi}+i\frac{4\pi}{e^2}\,,
\ee  
one sees immediately that 
\be \label{tautooneovertau}
\widetilde{\tau}=\frac{\widetilde\theta}{2\pi}+i\frac{4\pi}{\widetilde e^2}=-\frac 1{\tau}\,.
\ee 
In other words, the above procedure implements the symmetry transformation 
\be S: \tau\mapsto -\frac 1{\tau}\,,\ee 
which, along with the symmetry transformation $T: \tau\mapsto \tau+1$ that is induced by the periodicity of $\theta$, generate the S-duality group $SL(2;\Z)$.

\subsection{Multiple Maxwell fields and higher Buscher rules}
\label{sec32}

One may now ask what happens when multiple Maxwell fields are present in the theory. To this end, we consider $d$ 1-forms $A^{M}=A^{M}_{\m}(x)\dd x^{\m}$ for $M=1,\cdots,d$, essentially a $U(1)^{d}$ gauge theory. This is the direct analogue of the two-dimensional theory for multiple scalars. Motivated by this similarity, we consider a 
parallel of the string sigma model and write the following action in 4 dimensions: 
\be \label{SAM}
S[A^{M}]=-\frac 12 \int_{\S_4} \left(G_{MN} \, \dd A^{M}\w\ast\, \dd A^{N}+B_{MN} \, \dd A^{M}\w\dd A^{N}\right)\,.
\ee 
Multiple remarks are in order regarding this action and its comparison to the string sigma model. 
	The action \eqref{SAM} can be regarded as a sigma model for a map $A: T\S_4\to {\cal M}$ with components $A^{M}$, where $\S_4$ is the four-dimensional spacetime on which the theory lives and ${\cal M}$ is a suitable generalized target space.{\footnote{To provide a geometric justification of this statement, it would be useful to think of the spacetime and the target space as degree-shifted bundles $T[1]\S_4$ and $V[1]$ in the context of graded geometry. Recall that $T[1]\S_4$ is isomorphic to $T^{\ast}\S_4$ and thus degree-1 functions on it are ordinary 1-forms. Choosing degree-1 local coordinates $\xi^{M}$ on the target space, the fields $A^M$ are obtained as a pull-back under the ``sigma model'' map $A$, i.e. $A^M=A^{\ast}(\xi^M)$, see e.g. \cite{Gruetzmann:2014ica}. We will not elaborate further in this direction here.}} Unlike the scalar case, here $\cal M$ is not the arena of a physical theory; it is rather perceived as the usual target space of fields, as in every gauge theory. Note that in this interpretation we have made a slight abuse of notation, in that we identified the Lie algebra valued 1-form $A$ which lives in $\Omega^{1}(\S_4;\mf{u}(1)^d)$ with a map from $T\S_4$ to ${\cal M}$. 
	
	In the string sigma model, the couplings are functions of the scalar fields in general. Here one could think of a similar dependence for $G_{MN}$ and $B_{MN}$ or even couple the theory to additional scalars and allow the couplings to depend on them, a situation that is often encountered in supergravity theories. For the purposes of the present paper, we leave this as an open possibility since the following also hold when the couplings are field-independent in the above sense. We, however, mention the effect on the duality group due to the dependence of the couplings on dynamical fields when we discuss the duality groups in Section \ref{sec5}. Note that the single field case is simply obtained as the one with $d=1$ under the identification $G_{11}=1/e^2$. Electric/magnetic duality is then simply the map $G_{11} \to 1/16\pi^2G_{11}$ for the ``sigma model'' coupling. Not surprisingly, this is exactly the same behavior (up to the $16\pi^2$ factor) as in the scalar case. 
	
	As a final and important remark before we proceed, note that the ``$B$-field'' is now a symmetric tensor due to the different degree in this case. In fact, this stems from the existence of a gauge invariant $\theta$-term in Maxwell theory, which is absent for two-dimensional scalars but present here, as we discussed in the previous subsection. The off-diagonal part of the generalized $\theta$-term couplings $B_{MN}$ are absent in the single field case while the only remaining diagonal element is the usual $\theta$-term, i.e. for $d=1$, $B_{11}$ is identified with $\theta$ (up to physical constants).

	Bearing in mind the discussion below Eq. \eqref{SAM}, one may now follow the standard dualization procedure, essentially combining subsections \ref{sec22} and \ref{sec31}. We introduce again the 1-form $\widehat{A}$ and moreover we single out one of the Maxwell fields whose duality we will examine. We do this by splitting $A^{M}=(A^{m},A), m=1,\cdots,d-1$, where $A\equiv A^{d}$. We then rewrite the action \eqref{SAM} as 
	\bea
	S[A^{M}]&=&-\frac 12 \int_{\S_4}\left( G_{mn} \,\dd A^{m}\w\ast\,\dd A^{n}+2G_{md}\,\dd A^{m}\w\ast \,\dd A+G_{dd}\,\dd A\w \ast \,\dd A\right)  \nn\\ 
	&& -\, \frac 12 \int_{\S_4} \left(B_{mn}\,\dd A^{m}\w\dd A^{n}+2B_{md}\,\dd A^{m}\w\dd A+ B_{dd}\,\dd A\w \dd A\right)\,,\label{SAMsplit}
	\eea  
	always having in mind that $B$ is symmetric. Next, the corresponding first order action in terms of the independent 2-form $F$ is\footnote{We take the liberty to avoid introducing $4\pi$ as in \eqref{maxparent} in the Lagrange multiplier term, since this can be absorbed in the definition of $\widehat{A}\equiv\widehat{A}^d$. One should take this into account in order to make contact with the physical $\theta$ term in the single field case, but for the general purpose of this analysis it is an irrelevant redefinition.} 
	\bea 
	{\cal S}[A^{m},\widehat{A},F]&=&-\frac 12 \int_{\S_4}\left( G_{mn}\, \dd A^{m}\w\ast\,\dd A^{n}+2G_{md}\,\dd A^{m}\w\ast\, F+G_{dd}F\w \ast\, F\right)  \nn\\ 
	&& -\,\frac 12 \int_{\S_4}\left( B_{mn}\,\dd A^{m}\w\dd A^{n}+2B_{md}\dd A^{m}\w F+B_{dd}\,F\w F+2F\w\dd \widehat{A}\right)\,.\nn\\
	\label{SA1}
	\eea 
	Following the standard algorithm, the field equation for $\widehat{A}$ implies, at least locally, that $F=\dd A$. In other words, integrating out $\widehat{A}$ from the action \eqref{SA1} returns the original second order action \eqref{SAMsplit}. Now, the field equation obtained by variation of the 2-form $F$ is 
	 \be 
	B_{dd}\,F+G_{dd}\ast F+B_{md}\,\dd A^{m}+G_{md}\ast\dd A^{m}+\dd\widehat{A}=0\,,
	\ee  
which, as expected, contains both $F$ and $\ast F$. 
Its solution for $F$ is
\be 
F=u\,\dd\widehat{A}+v\ast\dd\widehat{A}+u_{m}\,\dd A^{m}+v_m\ast\dd A^{m}\,,
\ee 
with coefficients 
\bea 
&& u=-\frac{B_{dd}}{G_{dd}^2+B_{dd}^2}\,, \qquad v=\frac{G_{dd}}{G_{dd}^2+B_{dd}^2}\,,
\\[4pt]
&& u_m=-\frac{G_{md}G_{dd}+B_{md}B_{dd}}{G_{dd}^2+B_{dd}^2}\,, \quad v_{m}=\frac{B_{md}G_{dd}-G_{md}B_{dd}}{G_{dd}^2+B_{dd}^2}\,.
\eea 
	Inserting the resulting expressions into the first order action leads to the dual second order action 
	\be\label{eq:dual2ndorder} 
\widetilde{S}[\widehat{A}^{M}]	=-\frac 12\int_{\S_4} \left(\widetilde{G}_{MN}\,\dd \widehat{A}^{M}\w\ast\,\dd \widehat{A}^{N}+\widetilde{B}_{MN}\,\dd \widehat{A}^{M}\w\dd \widehat{A}^{N}\right)\,,
	\ee 
	where $\widehat{A}^{M}=(A^{m},\widehat{A})$ and the new couplings are given by the following 1-form Buscher rules:
	\bea 
	&&\widetilde{G}_{dd}=\frac{G_{dd}}{G_{dd}^2+B_{dd}^2}\,, \qquad \widetilde{B}_{dd}=-\frac {B_{dd}}{G_{dd}^2+B_{dd}^2}\,,\nn\\[4pt]
	&& \widetilde{G}_{md}=\frac{B_{md}G_{dd}-G_{md}B_{dd}}{G_{dd}^2+B_{dd}^2}\,, \qquad \widetilde{B}_{md}=-\frac{G_{md}G_{dd}+B_{md}B_{dd}}{G_{dd}^2+B_{dd}^2}\,,
	\nn\\[4pt]
	&& \widetilde{G}_{mn}=G_{mn}-\frac{G_{dd}\left(G_{md}G_{nd}-B_{md}B_{nd}\right)+B_{dd}\left(B_{md}G_{nd}+G_{md}B_{nd}\right)}{G_{dd}^2+B_{dd}^2}\,,\nn\\[4pt]
	&& \widetilde{B}_{mn}=B_{mn}-\frac{G_{dd}\left(G_{md}B_{nd}+B_{md}G_{nd}\right)-B_{dd}\left(G_{md}G_{nd}-B_{md}B_{nd}\right)}{G_{dd}^2+B_{dd}^2}\,.\label{1formBuscher}
	\eea 
	Note that when the diagonal part of $B_{MN}$ couplings in the duality direction vanishes ($B_{dd}=0$), one obtains a restricted set of rules that resembles more the standard stringy ones, where the $B_{MM}$ couplings are inadmissible anyway,
	\bea 
	&&\widetilde{G}_{dd}=\frac 1{ G_{dd}}\,,\qquad \widetilde{G}_{md}=\frac { B_{md}}{ G_{dd}}\,, \qquad \widetilde{B}_{md}=-\frac{G_{md}}{ G_{dd}}\,, \nn\\[4pt] 
	&& \widetilde{G}_{mn}=G_{mn}-\frac {G_{md}G_{nd}-B_{md}B_{nd}}{G_{dd}}\,,\qquad \widetilde{B}_{mn}=B_{mn}-\frac {G_{md}B_{nd}+B_{md}G_{nd}}{G_{dd}}\,. 
	\eea 
 Observe that although they are very similar to the standard T-duality rules \eqref{BuscherX}, they differ in a number of signs that reflect the change in degrees of the fields involved and in particular the fact that $B_{MN}$ is now a symmetric 2-tensor.  

 Furthermore, in the spirit of the $\tau$ parameter of the single field case, we can now define a complex metric 
\be 
\tau_{MN}=B_{MN}+i\,G_{MN}\,.
\ee 
It is then straightforward to cast the set of rules \eqref{1formBuscher} in a simpler form which directly generalizes \eqref{tautooneovertau}, namely
\be 
\widetilde{\tau}_{mn}=\tau_{mn}-\tau_{md}\,\frac 1{\tau_{dd}}\,\tau_{nd} \,,\quad \widetilde{\tau}_{md}=-{\tau}_{md}/{\tau}_{dd}\,,\quad \widetilde{\tau}_{dd}=-1/\tau_{dd}\,.
\ee 
 This already suggests the result for the transformation of the couplings when duality is performed along multiple directions in field space. Considering $A^M=(A^{m},A^{i})$, it reads as follows:
\be \label{buschergentau}
\widetilde{\tau}_{mn}= \tau_{mn}-\tau_{mi}{\tau^{ij}}\tau_{jn}\,, \quad \widetilde{\tau}_{mi}=-\tau_{mj}\tau^{ji}\,, \quad \widetilde{\tau}_{ij}=-\tau^{ij}\,,
\ee 
where $\tau^{ij}$ is the inverse of $\tau_{ij}$.
These expressions can be compared to \eqref{buschergenmetric}. We discuss this comparison in  Section \ref{sec5} in the context of twisted self duality for $p$-forms and mixed symmetry tensors of type $(p,1)$, after we first discuss the case of ordinary gravitons.

\section{Duality for gravitons in 4D}
\label{sec4}

In this section, we shift our attention to gravitons. Duality in the context of linearized general relativity has been studied extensively over the years, both in terms of field equations and of action functionals \cite{Hull:2001iu,West:2001as,Boulanger:2003vs}. Motivated by the results of the previous sections, our goal here is twofold. First, we want to examine this gravitational duality in the presence of $\theta$-terms, which was suggested at the on-shell level in \cite{Hull:2001iu}. We demonstrate that this can also be implemented off-shell and discuss what is the meaning of the $\theta$-term in that case. Second, we discuss the ``double'' duality of gravitons in the presence of $\theta$-term and comment on how the results of \cite{Henneaux:2019zod} can be modified in that case. In the spirit of completeness, we start with a short review of graded geometry and how Fierz-Pauli action looks like in this formulation. 

\subsection{Preliminaries on Fierz-Pauli and graded geometry}
\label{sec41}

Let us first briefly review some elementary facts related to the graded-geometric formulation of linearized gravity presented in \cite{Chatzistavrakidis:2016dnj,Chatzistavrakidis:2019len}, which will facilitate the analysis of duality. Recall that the Fierz-Pauli action for a massless graviton $h_{\mu\nu}$ in a flat 4D background reads as 
\be \label{FP}
S[h]=-\frac{M_{\text{P}}^2}{2g^2}\int \dd^4x\, \left(\partial^{\m}h^{\n\rho}\partial_{\mu}h_{\n\rho}-2\partial^{\mu}h^{\nu\rho}\partial_{\nu}h_{\mu\rho}+2\partial^{\mu}h^{\nu}{}_{\nu}\partial^{\rho}h_{\rho\mu}-\partial^{\mu}h^{\nu}{}_{\nu}\partial_{\mu}h^{\rho}{}_{\rho}\right),
\ee  
where $M_{\text{P}}$ and $g$ are the reduced Planck mass and a dimensionless reference constant respectively. The action is invariant under linearized diffeomorphisms with vector gauge parameter $\lambda_{\mu}$: 
\be 
\delta h_{\mu\nu}=\partial_{\mu}\l_{\nu}+\partial_{\nu}\lambda_{\mu}\,.
\ee
To avoid dealing with multiple terms that enter the action with different relative factors and index combinations, one can introduce a formalism similar to differential forms, but this time for mixed symmetry tensors like the graviton.{\footnote{A closely related approach in terms of ``multi-forms'' was employed in Refs. \cite{Bekaert:2002dt,deMedeiros:2002qpr,Bekaert:2003az,deMedeiros:2003osq}.}} To this end, let us consider a spacetime extended by two sets of anticommuting (degree-1) coordinates{\footnote{One should avoid confusing $\theta^{\mu}$ with the unrelated theta parameter $\theta$.}} $\theta^{\mu}$ and $\widetilde{\theta}^{\mu}$, which satisfy 
\be 
\{\theta^{\mu},\theta^{\nu}\}=0\,, \quad \{\widetilde{\theta}^{\mu},\widetilde{\theta}^{\nu}\}=0\,,\quad [\theta^{\mu},\widetilde{\theta}^{\nu}]=0\,.
\ee  
This spacetime corresponds to a graded supermanifold. 
Note the rather unconventional choice of sign when commuting coordinates across the two sets; $\theta^{\mu}$ and $\widetilde{\theta}^{\mu}$ commute.{\footnote{There is nothing deep about this convention and one could have as well chosen the more customary, opposite one, without changing the meaning of the results.}} Essentially, the two coordinates may be assigned degrees $(1,0)$ and $(0,1)$ respectively. Then any function on the graded supermanifold may be expanded in these coordinates with coefficients being tensor fields of certain type. Specifically, a function of given degree $(p,q)$, which is nothing but the number of $\theta$s and $\widetilde{\theta}$s in the expansion respectively, corresponds to a mixed symmetry tensor field of type $(p,q)$, i.e. a two-column Young tableaux representation of the general linear group. In the following, we will often use the words ``function'' and ``tensor'' interchangeably in the above sense. For instance, the components $h_{\mu\nu}$ of the graviton can be considered as the coefficients of the Taylor expansion of a degree-$(1,1)$ function{\footnote{The notation $h$ should not be confused with the trace of $h_{\mu\nu}$, which we always denote explicitly as $h^{\mu}{}_{\mu}$.}} $h$ in this extended space:
\be 
h=h_{(\mu\nu)}(x)\theta^{\mu}\widetilde{\theta}^{\nu}\,,
\ee
where symmetrization is considered with weight 1. The same is true for the Minkowski metric $\eta=\eta_{(\mu\nu)}\theta^{\mu}\widetilde{\theta}^{\nu}$ and any other mixed symmetry tensor. In addition, two (homological) vector fields may be defined as
\be 
\dd=\theta^{\mu}\frac{\partial}{\partial x^{\mu}} \quad \text{and} \quad \widetilde{\dd}=\widetilde{\theta}^{\mu}\frac{\partial}{\partial x^{\mu}}\,,
\ee 
both squaring to zero by definition, just like the ordinary exterior derivative of differential forms. At this stage, the only further ingredient we need to write the graded-geometric analog of the Fierz-Pauli action is a generalized Hodge star operator on the extended manifold, which will yield a suitable inner product. According to \cite{Chatzistavrakidis:2016dnj}, for any degree-$(p,q)$ function $\omega$ in $d$ dimensions, $\star\,\omega$ ,the generalized Hodge star of $\omega$, is a degree-$(d-p,d-q)$ function defined as 
\be 
\star\omega =\frac 1{(d-p-q)!}\,\eta^{d-p-q}\,\widetilde\omega\,
\ee 
where $\widetilde\omega$ is the degree-$(q,p)$ function obtained upon exchanging all $\theta$s with $\widetilde\theta$s in the Taylor expansion of the function $\omega$. Alternatively, it may be defined through the partial Hodge star operators $\ast$ and $\widetilde{\ast}$, which are simply the ordinary operators that in the present context yield a degree-$(d-p,q)$ and degree-$(p,d-q)$ function respectively, through the epsilon tensor. Then the generalized Hodge star $\star$ is given as \cite{Chatzistavrakidis:2019len}
	\be \label{starrelations}
\star \o=\ast\,\widetilde{\ast}\,(-1)^{\e(p,q)}\,\sum_{n=0}^{\text{min}(p,q)}\frac {(-1)^n}{(n!)^2}\,\eta^n\,\text{tr}^n\,\o~,
\ee  
where $\e(p,q)=(d-1)(p+q)+pq+1$ and one can define the trace in the present context as 
\be 
\text{tr}=\eta^{\mu\nu}\frac{\partial}{\partial\theta^{\mu}}\frac{\partial}{\partial\widetilde{\theta}^{\nu}}\,.
\ee Observe that $\star$ is not simply the product of the two ordinary operators, but rather contains traces of the mixed symmetry tensor, a feature that is crucial for actions such as \eqref{FP}.
 
  Using the above ingredients, the Fierz-Pauli action \eqref{FP} may be written in the compact form
\be \label{kineticgraded}
S[h]=-\frac{M_{\text{P}}^2}{4g^2} \int\dd^4 x\,\dd^4\theta\, \dd^4\widetilde{\theta}\,\dd h \star \dd h \equiv -\frac{M_{\text{P}}^2}{4g^2}  \int_{\hat{\S}_4}\, \dd h\star\dd h\,,
\ee 
where we introduced a shorthand notation that comprises the ordinary spacetime integral and the two Berezin integrals over the degree-1 coordinates $\theta$ and $\widetilde{\theta}$: 
\be 
\int_{\hat{\S}_4}:=\int \dd^4x\int_{\theta,\widetilde{\theta}} \equiv \int \dd^4x\,\dd^4\theta\,\dd^4\widetilde{\theta}\,,
\ee and similarly in arbitrary dimensions. To be more precise, \eqref{kineticgraded} leads to the Fierz-Pauli action with an additional total derivative term 
\be 
\frac{M_{\text{P}}^2}{2g^2}\int \dd^4x\,\partial_\m\left(h^{\m\n}\partial^\rho h_{\rho\n}-h_{\n\rho}\partial^{\n}h^{\m\rho}\right)\,.
\ee This should be kept in mind because we will deal with further total derivatives in the following. With a slight abuse of terminology, we will still refer to \eqref{kineticgraded} as the Fierz-Pauli Lagrangian. For completeness, note that the mass term in the Fierz-Pauli action for the massive graviton, $m^2\left(h^{\mu\nu}h_{\mu\nu}-h^{\mu}{}_{\mu}h^{\nu}{}_{\nu}\right)$ is simply $m^2\int_{\th,\widetilde{\theta}} h\star h$.

 The appeal of this formalism is twofold. First, it encodes the complicated structure of the Fierz-Pauli action in a single quadratic term. Second and more importantly, the formula \eqref{kineticgraded} holds for \emph{any} mixed symmetry tensor field of degree $(p,q)$ in \emph{any} dimension (upon promoting $\hat{\S}_4$ to $\hat{\S}_{\text{dim}\,\S}$ of course.) Thus it unifies the kinetic terms for all types of bosonic fields, be they scalars, $p$-forms or $(p,q)$ mixed symmetry tensors \cite{Chatzistavrakidis:2019len} (see also \cite{Chatzistavrakidis:2020gvy} for an alternative presentation highlighting this feature). 

\subsection{Off-shell duality for linearized gravity with theta term}
\label{sec42}

Having a degree-$(1,1)$ function $h$ with components being the symmetric tensor corresponding to the linearized graviton, one may use the two homological vector fields to define a degree-$(2,2)$ function $R$, whose components are the linearized Riemann tensor: 
\be 
R=\dd\widetilde\dd\,h\,,
\ee    
whereupon the linearized Einstein equations become 
\be \label{einsteineq1}
\text{tr}\,R=0\,.
\ee 
The corresponding Bianchi identities at the linearized level simply read as \be \label{BI1}\dd R=0 \quad \text{and} \quad \widetilde\dd R=0\,.\ee Gravitational duality in 4D is then the twisted self-duality based on 
\be \label{DR1}
\widehat R=\frac{1}{g^2}\ast R\,.
\ee
Recall that $\ast$ is the partial Hodge star operator that dualizes only the $\theta$-sector of $R$ and leaves the $\widetilde\theta$-sector untouched and therefore $\widehat R$ is a degree-$(2,2)$ function too. In the dual picture, $\widehat R=\dd\widetilde\dd\,\widehat h$ is the linearized Riemann tensor for a dual graviton $\widehat h$, which is also of degree-$(1,1)$ in 4D. This may be generalized to arbitrary spacetime dimension $\text{dim}\,\S$, where the dual graviton is of degree $(\text{dim}\,\S-3,1)$, but this is not important for our purposes here.

 One can further understand this duality as follows. In the original theory the Riemann tensor $R$ satisfies the field equations \eqref{einsteineq1} and the Bianchi identities \eqref{BI1}. What is more, $R$ is an irreducible tensor. This may be formalized in terms of a co-trace operator $\sigma$, whose action on $R$ vanishes 
\be 
\sigma\,R=0\,.
\ee
This operator, along with the corresponding dual one $\widetilde\s$, is defined in $\text{dim}\,\S$ spacetime dimensions as 
\be 
\s=(-1)^{1+(\text{dim}\,\S)(p+1)}\ast \text{tr} \ast\quad \text{resp.} \quad \widetilde\s=(-1)^{1+(\text{dim}\,\S)(p+1)}\,\widetilde\ast \,\text{tr}\, \widetilde\ast\,,
\ee 
and maps a function of degree $(p,q)$ to a function of degree $(p+1,q-1)$, respectively $(p-1,q+1)$. In the graded geometric formalism this simply means that 
\be 
\s=-\theta^{\mu}\frac{\partial}{\partial\widetilde{\theta}^{\mu}} \quad \text{and} \quad \widetilde{\s}=-\widetilde\theta^{\mu}\frac{\partial}{\partial\theta^{\mu}}\,.
\ee 
These operators offer a criterion on whether a tensor is irreducible \cite{Bekaert:2002dt,deMedeiros:2002qpr}. Specifically, a $(p,q)$ tensor $\omega$ is irreducible under the general linear group if and only if $\s\o=0$ for $p\ge q$ or $\widetilde\s\o=0$ for $p\le q$. For $p=q$ one should additionally require that $\omega=\widetilde\omega$. 
We can now readily see what the duality relation \eqref{DR1} is all about. The irreducibility of $R$ is mapped under the duality to the Einstein equation for the dual graviton, since $0=\s R\propto \s\ast\widehat{R} \propto \ast\, \text{tr} \widehat{R}$ and therefore $\text{tr}\widehat{R}=0$. Vice versa, the Einstein equation for $h$ is mapped to the irreducibility of $\widehat{R}$. Moreover the Bianchi identities for $\widehat{R}$ follow as well from the ones of $R$ and the Einstein equation. 
This on-shell duality may be implemented off-shell when a parent action which is first order in derivatives is considered \cite{West:2001as,Boulanger:2003vs}. 

However, it was suggested by Hull \cite{Hull:2001iu} that this twisted self-duality may be extended when one considers a gravitational theta angle aside the usual gravitational coupling constant.  The corresponding duality relation is 
\be \label{HullDR+}
 \widehat{R}=\frac 1{g^2}\ast R -\theta R\,,
\ee
where both $g$ and $\theta$ are dimensionless and $g$ may be absorbed in the gravitational coupling. This is then very similar to the $SL(2;\Z)$ duality for the 1-form in 4D. 
Our purpose in the ensuing is to implement this enhanced gravitational duality off-shell and to understand what is the theta term that accompanies the Fierz-Pauli action in that case. 
We work with Lagrangian densities instead of actions in this section; the action is simply obtained with the appropriate integration. Our starting point is the following parent Lagrangian:
\begin{equation}\label{parent2}
    \mf{L}[f,\l]=-\frac{M_{\text{P}}^2}{4}\int_{\theta,\widetilde\theta}\, f\star\left(\frac{1}{g^2}+\theta \,\ast\right)\mathcal{O}f-2\int_{\theta,\widetilde\theta}\,f\ast\widetilde{\ast}\,\dd^\dagger\lambda\,,
\end{equation}
where the (Berezin) integration is over the degree-1 coordinates only.
This Lagrangian contains two fields $f$ and $\l$, both being reducible mixed symmetry tensors of degrees $(2,1)$ and $(3,1)$ respectively. As in previous sections, the field $\l$ will act as a Lagrange multiplier for the Bianchi identity which will relate $f$ with the graviton, as we explain below. The operator $\dd^{\dagger}$ is simply the dual differential, defined as 
\be 
\dd^{\dagger}=(-1)^{1+(\text{dim}\,\S)(p+1)}\ast\dd\,\ast = \eta^{\mu\nu}\frac{\partial}{\partial\theta^{\mu}}\frac{\partial}{\partial x^{\nu}}\,,
\ee 
and similarly for $\widetilde{\dd}^{\dagger}$.
The last so far undefined ingredient of \eqref{parent2} is the algebraic operator $\mathcal{O}$. This was first defined in \cite{Chatzistavrakidis:2019len}, with the purpose of recovering a Lagrangian for an irreducible field when the Lagrange multiplier $\l$ is algebraically eliminated through its field equation. Although for differential forms this is automatic and thus ${\cal O}$ is the identity operator in that case, this is not true for mixed symmetry tensors. Indeed it was proven in \cite{Chatzistavrakidis:2019len} that for the graviton it takes the form 
\be\label{Ograviton} \mc O=\text{id}-\frac{1}{2}\widetilde{\sigma}\sigma\,,\ee  in terms of the co-trace operators $\s$ and $\widetilde{\s}$, where $\text{id}$ is the identity operator. The interested reader may check that the Lagrangian ${\mathfrak L}$ in components, obtained after performing the Berezin integrals, is 
\be 
\mf L[f,\l]=-\frac{M_{\text{P}}^2}{4g^2}\left(\frac 12f_{\m\n\rho}f^{\m\n\rho}+f_{\m\n\rho}f^{\rho\n\m}-2f_{\m\n}{}^\n f^{\m\rho}{}_{\rho}\right)-\frac{\theta M_{\text{P}}^2}{8}\vare^{\m\n\rho\s}f_{\m\n\k}f_{\rho\s}{}^\k-f_{\m\n\rho}\partial_\s \l^{\s\m\n\rho}.
\ee
Recall that $f_{\m\n\rho}$ is antisymmetric only in its first two indices. Although one could work with this component form, we find it less complicated to work directly with the graded formalism and present component expressions only when necessary for clarification.

Before we proceed, let us zoom in the theta term of the above Lagrangian. One may show, using that $f\star \ast\, \mc Of=-f\,\widetilde\ast f$, that it has the form 
\be \label{theta}
\mf L_{\theta}=\frac{\theta M_{\text{P}}^2}{4}\int_{\theta,\widetilde\theta}\, f\,\widetilde\ast \,f\,.
\ee This should not be surprising in view of the fact that the theta term in the 1-form case, which as stated earlier is very similar to the one we study here, is precisely of the form \eqref{theta} albeit for the 2-form field strength $F$ instead of the mixed symmetry tensor $f$. In other words, it is the analog of the second term of  \eqref{maxparent}, which may indeed be written in the form of \eqref{theta} in the graded formalism. We discuss the interpretation of this term at the end of this section. 

We now proceed with the investigation of the parent Lagrangian \eqref{parent2} in the same spirit as in previous sections, namely determining the second order Lagrangians that are obtained after  integrating out either the Lagrange multiplier $\l$ or the field $f$ through their field equations. In the process of doing that, we will also see how the duality relation \eqref{HullDR+} is implemented in this picture.  
The equation of motion for the field $\lambda$ is simply $\dd f=0$. This is a Bianchi identity for the field $f$, which can be solved locally as $f=\dd e$ for a \emph{reducible} tensor $e$ of type $(1,1)$. Note that the reducibility of this tensor is a feature that we did not encounter at all in the previous sections, however it is standard in dualities that involve mixed symmetry tensor fields. We can decompose such a tensor into symmetric and antisymmetric parts or, equivalently, 
\be\label{identification 1} 
e:=h+\widetilde{\sigma}\,b\,,
\ee 
where $h$ is the irreducible part of $e$ (which is going to be identified with the symmetric linearized graviton) and $b$ is the $2$-form component of $e$.{\footnote{Here we only use a 2-form in $\theta$s, namely a degree-$(2,0)$ function and as such $\widetilde{\s}b$ is a degree-$(1,1)$ function, as it should. One could also include a term $\s\widetilde{b}$, with a degree-$(0,2)$ function $\widetilde{b}$, but this is redundant and we ignore it in the following.}} Integrating out $\lambda$ leads to the second-order Lagrangian
\be \label{L_I}
\mathcal{L}[h,b]=-\frac{M_{\text{P}}^2}{4g^2}\int_{\theta,\widetilde\theta}\dd h\star\dd h+\mc L_{\partial}^{(g^2)}+\mc L_{\partial}^{(\theta)}\,,
\ee
where the first term is the Fierz-Pauli Lagrangian for $h$ and the last two terms correspond to total derivative contributions, specifically
\begin{equation}\label{BT I}\begin{split}
    &\frac{4g^2}{M_{\text{P}}^2}\mc L_{\partial}^{(g^2)}=\int_{\theta,\widetilde\th}\dd h\star \widetilde\dd b-\int_{\theta,\widetilde\th}\dd h\star \dd\widetilde{\sigma}b+\int_{\theta,\widetilde\th}\dd\widetilde{\sigma}b\star \widetilde\dd b\,,\\[4pt]
    &\frac{4}{\theta M_{\text{P}}^2}\mc L_{\partial}^{(\th)}=\int_{\theta,\widetilde{\theta}}\dd h\,\widetilde\ast\,\dd h+2\int_{\theta,\widetilde{\theta}}\dd h\,\widetilde\ast\,\dd \widetilde{\s}b+\int_{\theta,\widetilde{\theta}}\dd \widetilde{\s}b\,\widetilde\ast\,\dd \widetilde{\s}b\,.
    \end{split}
\end{equation}
Note that such terms are typically ignored, although they are relevant in pinpointing the symmetries of the theory, here in particular the local Lorentz symmetry. Here we carefully identify them because of the theta term in the action and we trace the duality also for them. It is useful at this stage to present the corresponding component expressions for clarity:
\bea 
\mc L_{\partial}^{(g^2)}&=& \frac{M_{\text{P}}^2}{2g^2}\,\partial_\m\left(
2h\,\partial_\n b^{\n\m}+2h_{\n\rho}\partial^\n b^{\m\rho}-b_{\n\rho}\partial^\n b^{\m\rho}+b^{\m\n}\partial^{\rho} b_{\rho\n}\right)\,,
\\[4pt]
\mc L_{\partial}^{(\th)}&=&-\frac{\theta M_{\text{P}}^2}{2}\, \partial_\m\left(\vare^{\m\n\rho\s}h_{\n\k}\partial_{\rho} h_\s^\k+2\vare^{\m\n\rho\s}h_{\n\k}\partial_{\rho} b_\s{}^\k+\vare^{\m\n\rho\s}b_{\n\k}\partial_{\rho} b_\s{}^\k\right)\,.
\eea 
 
In the second leg of the duality, we compute the Euler-Lagrange equations for the field $f$, which read as 
\begin{equation}\label{duality relation}
    \star\left(\frac{1}{g^2}+\theta \,\ast\right)\mathcal{O}f=-\frac{4}{M_{\text{P}}^2}\ast\widetilde\ast \,\dd^\dagger\lambda\,.
\end{equation}
This is a duality relation. We can bring it in a more illuminating form as follows. 
Taking into account the relation of $\star$ to the partial Hodge operators laid out in \eqref{starrelations}, it may be rewritten as
\begin{equation} \label{9}
(1-\eta\,\text{tr})\left(\frac{1}{g^2}+\theta \,\ast\right)\mathcal{O}f=-\frac{4}{M_{\text{P}}^2}\dd^\dagger\lambda\,.
\end{equation}
Taking the trace on both sides, we obtain: 
$$\text{tr}\left[\left(\frac{1}{g^2}+\theta \,\ast\right)\mathcal{O}f\right]=\frac{2}{M_{\text{P}}^2}\text{tr}\,\dd^\dagger\lambda$$
This can be substituted back into \eqref{9} to give:
\begin{equation} \label{DR}
\left(\frac{1}{g^2}+\theta \,\ast\right)\mathcal{O}f=-\frac{4}{M_{\text{P}}^2}\left(1-\frac{1}{2}\eta\,\text{tr}\right)\dd^\dagger\lambda\,.
\end{equation}
This is then the final form of the duality relation that results directly from the parent Lagrangian as an equation of motion for the field $f$. At this stage, we would like to compare it to the on-shell duality relation \eqref{HullDR+}. This can easily be done by using the explicit form of the operator $\mc O$ in terms of the co-trace maps and collapsing $f$ to its on-shell expression $f=\dd(h+\widetilde{\s}b)$. Moreover, at the interest of identifying the dual graviton $\widehat{h}$, the Lagrange multiplier $\l$ may be decomposed as 
\be\label{decomposition} \lambda=\widehat{\lambda}+\eta\mathring{\lambda}\,,\ee
where $\widehat\lambda$ and $\mathring{\lambda}$ are the traceless and trace parts of $\lambda$, of respective degrees $(3,1)$ and $(2,0)$. Furthermore, we make the following identifications
\be\label{identification 2}
\widehat\lambda:=-\frac{M_{\text{P}}^2}{4}\ast\widehat h\,,\qquad \mathring{\lambda}:=-\frac{M_{\text{P}}^2}{4} \ast\widehat b\,,
\ee
where we introduced the dual graviton $\widehat h$ and a new $2$-form $\widehat{b}$. Note that $\widehat{h}$ is irreducible by definition, as the Hodge dual of a traceless tensor. Then, a simple algebraic calculation that we present in  Appendix \ref{appa} shows that the duality relation \eqref{DR} becomes on-shell 
\be \label{onshellDR}
\left(\frac{1}{g^2}+\theta \,\ast\right)(\dd h-\widetilde{\dd}\,b)=-\ast \dd\widehat{h}+\ast\,\widetilde{\dd}\,\widehat{b}\,.
\ee
Acting on both sides of this equation with the combined operator  $\ast\,\widetilde{\dd}$  and defining the irreducible Riemann tensors $R:=\dd\,\widetilde{\dd}\,h$ and $\widehat{R}:=\dd\,\widetilde{\dd}\,\widehat h$ as before, one can see that the $2$-forms $b$ and $\widehat{b}$ decouple due to the nilpotency of $\widetilde{\dd}$. The resulting relation reads as
\begin{equation}\label{Hull++}
\widehat R=\frac{1}{g^2}\ast R-\theta R\,,
\end{equation}
and is precisely the duality relation \eqref{HullDR+} relating the Riemann tensor with its dual at the on-shell level. 

The next and final step of the procedure is to determine the dual Lagrangian in terms of the dual fields $\widehat{h}$ and $\widehat{b}$. To do so, it is necessary to solve the duality relation and determine $f$. This is slightly more complicated than the scalar or $p$-form case, and we present the details in Appendix \ref{appa}. The result is   
\be \label{solution DR}
f=-\frac{g^2}{1+\theta^2 g^4}(1-\widetilde\sigma\sigma)(1-\theta g^2\ast)\ast (\dd\widehat h-\widetilde{\dd}\,\widehat{b}).
\ee
The dual Lagrangian can now be determined by substituting the duality relation \eqref{duality relation} and its solution \eqref{solution DR} back into the parent Lagrangian \eqref{parent2}. Some basic algebraic manipulations, also presented in detail in Appendix \ref{appa}, lead to the final second order dual Lagrangian
\be \label{L_II}
\widetilde{\mathcal{L}}[\widehat{h},\widehat{b}]=-\frac{M_{\text{P}}^2}{4\widetilde{g}^2}\int_{\theta,\widetilde{\theta}}\dd \widehat h\star\dd \widehat h+\widetilde{\mc L}_{\partial}^{(\widetilde g^2)}+\widetilde{\mc L}_{\partial}^{(\widetilde\theta)}\,,
\ee
where the first term is the Fierz-Pauli Lagrangian for the dual graviton $\widehat h$ and the last two terms are the same total derivative contributions as in Eqs. \eqref{BT I}, with fields replaced by their hatted counterparts $\widehat{h}$ and $\widehat{b}$. This is then the complete picture of how the duality \eqref{HullDR+} is implemented off-shell. Naturally, the new couplings $\widetilde g$ and $\widetilde \theta$ are related to the original ones through
\be \label{couplings trafo}
\frac{1}{\widetilde g^2}=\frac{g^2}{1+\theta^2g^4}\,,\qquad \widetilde\theta= -\frac{\theta g^4}{1+\theta^2g^4}\,.
\ee
Defining the complexified coupling $\tau:=\theta+\frac{i}{g^2}$, these transformations are equivalent to the inversion rule
$
\tau\mapsto -\frac{1}{\tau}\,,
$
as described at the on-shell level in Ref. \cite{Hull:2001iu}. 

As for scalars and $1$-forms, the discussion in the present section may be generalized in two directions. The first is to consider multiple gravitons $h^{M}, M=1,\dots,d\,,$ and the second is to consider generalized mixed symmetry tensor fields of degree $(p,1)$. To avoid repetition, we discuss these cases collectively, along with $p$-forms, in  Section \ref{sec5}. 

Finally, let us return to the discussion of the theta term that accompanies the linearized gravity action in the two dual second-order actions \eqref{L_I} and \eqref{L_II}. Ignoring for the moment terms that contain the 2-form $b$, we observe that there is a quadratic term in $h$ proportional to 
$\epsilon^{\m\n\rho\s}\partial_{\mu}h_{\n\k}\partial_{\rho}h^{\k}_{\s}$, which in our notation is simply the term $\dd h\,\widetilde\ast\,\dd h$ under the Berezin integral. This contains only first derivatives of $h_{\m\n}$ and moreover it involves an epsilon tensor. As such it should be associated to a topological invariant. However, it does not arise in the linearized limit of the Pontryagin density, which involves the Riemann tensor instead and therefore second derivatives of $h_{\m\n}$. As shown in \cite{Chatzistavrakidis:2020wum}, it is the linearized version of a topological invariant in a spacetime with torsion equipped with the Weitzenb\"ock connection. Invariants formed through the torsion tensor are the so-called Nieh-Yan terms \cite{Nieh:1981ww,Chandia:1997hu,Li:1999ue} and hence the term we encounter here is a special case of them. This shows that Hull's duality relation \eqref{HullDR+} is realized for such spacetimes only. 

It is useful to trace the relation of this gravitational theta term to the theta term in electromagnetism. The latter is simply the product of the electric and magnetic fields, $\vec E \cdot \vec B$ in the Lagrangian. The former is very similar, since one may decompose the tensor $h_{\m\n}$ in gravitoelectric (Newtonian potential) $\phi$ and gravitomagnetic $A_i$ potentials as follows 
\be 
\phi=-\frac{c^2}{8}h_{00}\,,\qquad A_i=\frac{c^2}{2}h_{i0}\,.
\ee
Defining the gravitoelectric and gravitomagnetic fields as 
\be 
\vec{E}_{g}=-\vec{\nabla}\phi\,,\qquad \vec{B}_{g}=\vec{\nabla}\times \vec{A}\,,
\ee  
the gravitational theta term in the action turns out to be proportional to their product $\vec{E}_{g}\cdot\vec{B}_{g}$. This yields a modified version of gravitoelectromagnetism \cite{Mashhoon:2003ax}. It becomes particularly interesting in situations where $\theta$ is not the same in all regions of spectime or when it is promoted to a background field, resulting in a gravitational analogue of axion electrodynamics. This viewpoint can lead to topological magnetoelectric effects, similar to topological insulators, as suggested in \cite{Chatzistavrakidis:2020wum}. It is also worth noting that $\theta$ could be promoted to a dynamical axion-like field in nonlinear gravity with torsion, in which case it can have important physical implications e.g. in addressing the strong CP problem \cite{Mercuri:2009zi,Lattanzi:2009mg,Castillo-Felisola:2015ema}.

\subsection{Double duality with theta term}

Apart from the gravitational duality that we dealt with in section \ref{sec42}, there is a further possibility that arises; one can consider dualizing both slots of the $(1,1)$ tensor $h$, using $\ast$ and $\widetilde{\ast}$ \cite{Hull:2001iu}. This is referred in the literature as double duality and in $\text{dim}\,\S$ dimensions it leads to a $(\text{dim}\,\S-3,\text{dim}\,\S-3)$ dual mixed symmetry tensor field ${\widecheck{h}}$. Here we continue working in 4D. In the presence of a theta term, it was argued that the on-shell duality relation in this case is \cite{Hull:2001iu} 
\be \label{HullDR2+}
 {\widecheck{R}}=-\frac{2\theta}{g^2}\ast R+\left(\theta^2-\frac{1}{g^4}\right)R\,.
\ee 
We observe that in the absence of $\theta$, one obtains that ${\widecheck{R}}\propto R$ and therefore the two dual gravitons are related algebraically and not through a Hodge duality relation, as also observed and discussed recently in \cite{Henneaux:2019zod} for the 5D case. However, this is not the case for non-vanishing theta angle. 

Our purpose here is to explain how this duality relation is obtained from the parent action, as in the previous subsection. Note that the off-shell dualization to the double dual graviton was first considered in \cite{Boulanger:2012df}. It is also implemented in the present graded formalism in \cite{Chatzistavrakidis:2019len}.  Here we go one step further and include the theta term. Fortunately, not many changes are required in this context and the parent action is essentially the same as the one we considered in the previous subsection.

The double dual graviton ${\widecheck h}$ results from dualizing $\widehat{h}$, in exactly the same way that $\widehat{h}$ results from dualizing $h$. 
This double dualization procedure follows the exact same steps as in section \ref{sec42}, with the same parent action. It is  obtained by appropriately modifying the identifications \eqref{identification 1} and \eqref{identification 2} into 
\be \label{identification 3}
e:=\widehat{h}+\widetilde{\s}\,\widehat{b}\qquad\text{and}\qquad \widehat{\l}:=-\frac{M^2_{\text{P}}}{4}\ast {\widecheck h}\,,\quad \mathring{\l}:=-\frac{M^2_{\text{P}}}{4}\ast {\widecheck b}\,,
\ee
respectively, such that the original second order action is nothing but the (previously dual) action \eqref{L_II}. Therefore, carrying out the same procedure, the on-shell duality relation turns out to be 
\be 
{\widecheck R}=\frac{1}{g^2}\ast \widehat{R}-\theta\widehat{R}
\ee
between the dual and double dual Riemann tensors $\widehat R=\dd\widetilde\dd\,\widehat h$ and $ {\widecheck R}=\dd\widetilde\dd\,  {\widecheck h}$. This is formally the same as the duality relation \eqref{Hull++} between the original and the dual Riemann tensor. Using this, one can rewrite the above equation as
\be \label{HULL2}
{\widecheck R}=-\frac{2\theta}{g^2}\ast R+\left(\theta^2-\frac{1}{g^4}\right)R\,,
\ee
which is precisely \eqref{HullDR2+}. The dual action  becomes the expected second order one for the double dual graviton, as before. It is worth noting though that this is the case only in 4D, which is our working dimension here. In higher dimensions, the action obtained for the double dual graviton using this procedure would contain additional off-shell fields that cannot be eliminated algebraically \cite{Boulanger:2012df}. These do not arise in 4D though.

\section{Duality for $p$\,-forms and $(p,1)$ tensor fields in ($2p+2$)D}   
\label{sec5}

Our detailed analysis up to this point was limited to single and multiple scalars or 1-forms and to a single graviton. However, this can be readily generalized for single or multiple $p$-forms, multiple gravitons and tensor fields of degree $(p,1)$.{\footnote{The study of such fields was initiated by Curtright \cite{Curtright:1980yk}. Among other uses they appear in exotic maximal supergravity theories in 6D \cite{Hull:2000rr}, for example ${\cal N}=(3,1)$ theory whose action was constructed in recent years \cite{Henneaux:2018rub,Bertrand:2020nob}. In that context they are sometimes called ``exotic gravitons''.}} In this section, we generalize the discussions of previous sections to all these cases in dimensions where (twisted) self-duality\footnote{It is important to note that we are interested in exploring the duality in the presence of generalized $\theta$-term where the notion of (twisted) self-duality is no longer strictly applicable. To be precise, what we do is to include the $\theta$-term in those dimensions where in the absence of $\theta$-term (twisted) self-duality is possible.} is possible. This condition immediately implies that the dimension of spacetime (the ``source'' $\Sigma$ in the generalized sigma model perspective) must be $\text{dim}\,\S=2p+2$. Henceforth we must distinguish $p=2k$ and $p=2k-1$, $k\in\N$, in which cases $\text{dim}\,\S$ is $4k+2$ and $4k$ respectively. 
This restriction on the dimension holds for both $p$-forms and $(p,1)$ mixed symmetry tensors. This is because in the latter case we are essentially performing dualization in the first slot, with the exception of the special case of $(1,1)$ where we also discussed double duality. For instance one can see this easily for the graviton in 4D from \eqref{DR1}, where the dual curvature is related to the curvature by the use of the Hodge duality operator $\ast$, which only dualizes the first slot. 
This allows us to treat the duality of multiple $p$-forms or $(p,1)$ mixed symmetry tensors in $2p+2$ dimensions on equal footing. 

\subsection{Charting the duality rules for  $p$-forms and $(p,1)$-gravitons}\label{sec51}

In our approach, the starting point for the dualization of either multiple $p$-forms or $(p,1)$ tensor fields is universal and therefore the corresponding generalization of the analysis of previous sections is fairly straightforward. The universal parent action that effectuates the dualization procedure along a single direction is 
\be\label{Suniv} 
\mc S[F^M,\Lambda]=\int \dd^{2p+2}x\, \mf{L}[F^M,\Lambda]\,,
\ee 
with the parent Lagrangian density being 
\begin{equation}\label{universal}
\mathfrak{L}[F^M_{p+1,q},\Lambda_{p+2,q}]= (-1)^{p+q} \alpha^2 \left[ \int_{\theta,\widetilde{\theta}} F^M \star \mathcal{U}_{MN} \mathcal{O} F^N  -2 \int_{\theta,\widetilde{\theta}} F\ast\widetilde{\ast}\,\dd^\dagger\Lambda\right]\,, \quad q=0,1\,,
\end{equation}
where
\be
\mathcal{U}_{MN} = G_{MN}+ (-1)^{p+q} B_{MN}\,\ast \, .
\ee
This is a functional of two fields $F^M$, $M=1,\dots d$ and $\Lambda$, in general reducible under the general linear group,\footnote{In order to make a connection with the action of a single graviton discussed earlier, one can set $\l = \alpha^2 \Lambda$. Using $\Lambda$ allows us to put the action for all types of $(p,q)$ fields in the same footing without introducing an additional dimensionful constant. Also, we work in the natural unit ($M_p = 1$) and $\alpha\in \mathbb{C}$.} with degrees as indicated in \eqref{universal}. The field $F^{M}$ is decomposed as $F^{M}=(F^{m}=\dd A^m,F)$, where $m=1\dots, d-1$ and $F\equiv F^{d}$, with $A^{m}$ being $d-1$ potentials of degree $(p,q)$ and $\Lambda = \ast \widehat{e}$ with $\widehat{e}$ being of degree $(p,q)$. Recall that $\widehat{e}=\widehat{h}+\widetilde{\sigma}\widehat{b}$ in the case of a single graviton of Section \ref{sec4}. In the present context, $q$ takes the values 0 or 1, the first being relevant for the dualization of $p$-forms and the second for $(p,1)$ tensors. The operator ${\cal O}$ depends on $p$ and $q$, being given by 
\be 
\mc O= \begin{cases}\text{id}\,, \quad & q=0\,, p\ge 0 \\ \text{id}-\frac 1{p+1}\widetilde\s\s\,, & q=1\,,p\ge 1\end{cases}\,,
\ee     
as found in \cite{Chatzistavrakidis:2019len} for the single field case. Note that since $\mc O$ is the identity for $q=0$, and $F^{M}$ and $\Lambda$ do not contain any $\widetilde{\theta}$ coordinates in that case, the Berezin integral over $\widetilde{\theta}$ can be performed explicitly and the parent action for $p$-forms is reduced in ordinary notation to 
\be 
{\mc S}[F_{p+1}^M,\Lambda_{p+2}]=-\alpha^2 \int_{\Sigma_{2p+2}} \left[ F^{M}\w \ast \left(G_{MN} + (-1)^{p} B_{MN} \ast \right) F^{N} +2\,(-1)^{p+1}  F \w \dd \widehat{A} \right]\,,
\ee 
with $\a=1/\sqrt{2}$ to match the convention used in Sections \ref{sec2} and \ref{sec3}.
Finally,  with reference to \eqref{universal}, although $G_{MN}$ is always symmetric, the symmetry properties of $B_{MN}$ evidently depend on the degree $p$, being symmetric for odd $p$ and antisymmetric for even $p$. This completes the characterization of the universal parent Lagrangian \eqref{universal}.

  Following the procedure described in all previous sections, one easily finds out that the action \eqref{Suniv} gives rise to two second order actions for two dual sets of $(p,q)$ fields $A^{M}=(A^m,A)$ and $\widehat{A}^{M}=(A^m,\widehat{A})$, both of degree $(p,q)$ with $q=0,1$: 
  \be 
  S[A^M] \, \xleftarrow{\,\Lambda\, \, \text{on-shell}} \, {\cal S}[F^M,\Lambda] \, \xrightarrow{\,F\, \, \text{on-shell}} \, \widetilde{S}[\widehat{A}^M]\,,
  \ee 
with $\dd A^M=F^M$ and $\widehat{A}$ defined via $\Lambda=\ast\widehat{A}$. 
For instance, to match our conventions, $\a^2=-1/4$ for the graviton, i.e. $(p,q)=(1,1)$, where $\widehat{A}=\widehat h+\widetilde\s \widehat b$. Note that although in the $p$-form case $\Lambda$ is traceless, in the $(p,1)$ case both actions $S$ and $\widetilde{S}$ depend also on the trace part of $\Lambda$ through total derivative terms, as discussed in Section \ref{sec42} for the graviton. Evidently, this procedure is associated to a duality relation, encompassing all cases discussed previously as well as generalizations thereof. For example, one finds an on-shell duality relation for multiple gravitons, 
  \be \label{DRmulti}
   \widehat R= G_{dd}\ast R-B_{dd}R+G_{dm}\ast R^{m}-B_{dm}R^{m}\,,
  \ee
  where $R^m=\dd\widetilde{\dd}h^m$ is the linearized curvature tensor for the spectator gravitons $h^m$. 
  Evidently, Eq. \eqref{DRmulti} extends \eqref{HullDR+} and reduces to it in the single field case where $G_{dd}=1/g^2$, $B_{dd}=\theta$ and the components $G_{dm}$ and $B_{dm}$ vanish. A similar extension holds for the second duality relation \eqref{HullDR2+}. Eq. \eqref{DRmulti} is then the direct analogue of Eq. \eqref{DRmultiscalar} for multiple scalars with $F=\dd X$. We note in passing that there are also equations of motion for the spectator fields in both cases, which should eventually be considered in addition to these multi-field duality relations.  
  
  The background fields of the two dual actions are related via either of the two sets of Buscher rules 
 \eqref{buschergenmetric} or \eqref{buschergentau} (in the present case of a single duality for $i,j=d$), depending on the value of $p$. For even $p=2k$, one obtains the rules \eqref{buschergenmetric}, whereas for odd $p=2k-1$ one obtains the alternative set \eqref{buschergentau}. We further discuss the meaning of these sets collectively in Section \ref{sec52}, with respect to  the group-theoretical underlying structure of the duality group for odd and even $p$. The results up to this point are summarized in Table \ref{Tab1}.
   
\begin{table}[t]
	\begin{center}
		\begin{tabular}{ |c|c|c|c|c| } 
			\hline
			{$(p,q)$ field } & $\mc E_{MN}$ & Buscher Rules & Examples \\
			\hline\hline
			{$p=2k-1$}  & $\tau_{MN}$ & Equation \eqref{buschergentau}  & 1-forms, gravitons
			  \\
			\hline\hline
			{$p=2k$} & $E_{MN}$ & Equation \eqref{buschergenmetric} & scalars, 2-forms, $(2,1)$ Curtright fields
			\\ 
			\hline
		\end{tabular}
		\caption{ Background field matrix $\mc E_{MN}$, set of Buscher rules for the background fields and basic examples for the dualization of $(p,q)$ fields with $q=0,1$ in $2p+2$ dimensions.}\label{Tab1}
	\end{center}
\end{table}

\subsection{Orthogonal vs. symplectic duality groups}
\label{sec52}

In this final part, we complete our discussion of the higher Buscher rules with a brief account on the group structure underlying the two sets of transformations which govern the behavior of background fields under duality for multiple $p$-forms or $(p,1)$ tensor fields, namely \eqref{buschergenmetric} for even $p$ and \eqref{buschergentau} for odd $p$. As expected, the first set is associated to orthogonal transformations in $O(d,d;\R)$ and the second set to symplectic transformations in $Sp(2d;\R)$, the latter as described in \cite{Gaillard:1981rj}. 

In the present context, we can discuss both cases in a parallel fashion. In the following, we assume there are $s$ Killing directions in the total $d$-dimensional target space. We first consider a $d\times d$ block matrix 
\be \label{edecomposition}
\mc E = \begin{pmatrix} \mc E_1 & \mc E_2 \\ \mc E_3 & \mc E_4 \end{pmatrix}\,,
\ee
with block sizes $s\times s$ for $\mc E_1$, $s\times (d-s)$ for $\mc E_2$, $(d-s)\times s$ for $\mc{E}_3$ and $(d-s)\times (d-s)$ for $\mc E_4$. In the two separate classes of cases we have discussed in this paper, it is identified as 
\be 
\mc E= \begin{cases} E\,, \quad  & p=2k\\ \tau\,, \quad  & p=2k-1\end{cases}\,.
\ee 
This means that $\mc E_{ij}=(\mc E_1)_{ij}$ is $E_{ij}$ or $\tau_{ij}$, ${\mc E}_{im}=({\mc E}_{2})_{im}$ is $E_{im}$ or $\tau_{im}$ etc.

Next, closely following \cite{Giveon:1994fu,Giveon:1991jj}, we consider block matrices 
\be 
g=\begin{pmatrix} a & b \\ c & d \end{pmatrix} \in \mc G\,,
\ee 
where $\mc G$ is either $O(s,s;\R)$ for $p=2k$ or $Sp(2s;\R)$ for $p=2k-1$. This means that they satisfy the condition 
\be \label{invariance}
g^{t} \mc J g=\mc J\,,
\ee 
for a constant $2s\times 2s$ matrix 
\be \label{calj}
\mc J=\begin{cases} J=\begin{pmatrix} 0 & \one_{s} \\ \one_{s} & 0\end{pmatrix}\,, &  p=2k \\[15pt] \Omega= \begin{pmatrix} 0 & \one_s \\ -\one_s & 0 \end{pmatrix}\,, & p=2k-1 \end{cases}\,.
\ee 
These can be embedded in the larger groups $O(d,d;\R)$ and $Sp(2d;\R)$, collectively denoted by $\widehat{\mc G}$, by means of elements 
\be 
\widehat{g}=\begin{pmatrix} \widehat{a} & \widehat{b} \\ \widehat{c} & \widehat{d} \end{pmatrix} \in \widehat{\cal G}\,,
\ee 
with blocks given as 
\be 
\widehat a=\begin{pmatrix} a & 0 \\ 0 & \one_{d-s} \end{pmatrix}\,, \quad \widehat b=\begin{pmatrix} b & 0 \\ 0 & 0 \end{pmatrix}\,, \quad \widehat c=\begin{pmatrix} c & 0 \\ 0 & 0 \end{pmatrix}\,, \quad \widehat d=\begin{pmatrix} d & 0 \\ 0 & \one_{d-s} \end{pmatrix}\,.
\ee 
Clearly $\widehat{g}$ satisfies \eqref{invariance} for the higher-dimensional $\widehat{\mc J}$, given by \eqref{calj} with $d\times d$ blocks instead.

The two sets of equations \eqref{buschergenmetric} and \eqref{buschergentau} are now associated to fractional linear transformations of $\mc E$ as 
\be 
\widetilde{\mc E}=\widehat{g}(\mc E)= (\widehat a\mc E+\widehat b)(\widehat c\mc E+\widehat d)^{-1}\,.
\ee  
This may be seen from the fact that when $\widetilde{\mc E}$ is decomposed according to \eqref{edecomposition}, one finds 
\begin{subequations} \bea 
\widetilde{\mc E_{1}}&=&(a \mc E_1+ b)(c\mc E_1+d)^{-1}\,, \\
\widetilde{\mc E_{2}}&=& (a-\widetilde{\mc E}_{1} c)\mc E_2\,, \\ 
\widetilde{\mc E}_3&=& \mc E_3(c \mc E_1+d)^{-1}\,, \\ 
\widetilde{\mc E}_4&=&\mc E_4-\mc E_3(c \mc E_1+d)^{-1}c\mc E_2\,.
\eea \end{subequations}
It is then straightforward to see that the transformations \eqref{buschergenmetric} and \eqref{buschergentau} are identified with elements of $\mc G$ corresponding to factorized dualities, namely 
\be 
g=\begin{pmatrix} \one-e_i & e_i \\ (-1)^{p}e_i & \one-e_1\end{pmatrix}\,,
\ee 
where $e_i=\text{diag}(0,\dots,0,1_i,0,\dots,0)$ and $1_i$ denotes that the entry is in the $i$-th position. For scalars where $p=0$ this is the usual story of factorized T-dualities \cite{Giveon:1994fu,Giveon:1991jj}, where the discrete T-duality group is $O(s,s;\Z)${\footnote{Note that for the scalar case its continuous version also maps between exact conformal backgrounds.}}. In this paper we have discussed that this generalizes to all $p$, both for multiple $p$-forms and for multiple $(p,1)$ generalized gravitons.  

As a final remark, we would like to draw attention to a subtlety that we have kept under the rug. Throughout the paper, we considered couplings $G_{MN}$ and $B_{MN}$ to be, in general, functions of scalar fields. However, we did not discuss the dynamics of those scalar fields. In such a case, we can treat the scalar fields as background fields therefore we are able to eliminate them by means of their equations of motion. These steps lead to a re-writing of the action with the couplings that are scalar independent, or in other words constant couplings. Such cases have been discussed in \cite{Gaillard:1981rj} where the duality group in the absence of scalars is a compact group. In fact, in the cases discussed in this paper, in the absence of scalars the duality group is the maximal compact subgroup of the relevant non-compact duality group in the presence of scalars \cite{Bunster:2011aw}, i.e. $U(1)^d$ (maximal compact subgroup of $Sp(2d,\mathbb{R})$) for odd $p$ and $O(d)$ (maximal compact subgroup of $O(d,d)$) for even $p$. In the presence of scalars, one requires that they transform in an appropriate way in order to keep the full Lagrangian (including the scalar Lagrangian) invariant. This lifts the compact duality group to the non-compact ones. In the above discussion, in a sense we are considering that the scalar fields are dynamical even though we are not explicitly writing down the scalar Lagrangian.

\section{Conclusions}
\label{sec6}

Duality is a key property of many physical theories that allows to probe the same physical effects with different degrees of freedom or observables. For instance, T-duality of string theory exchanges momentum and winding modes, whereas electric/magnetic duality exchanges the description of  electromagnetic fields in terms of electric or magnetic gauge potentials giving rise to dual observables such as Wilson and 't Hooft loops. Dualities also act on the couplings of a given theory in a very particular way. In the case of 2D compact scalars, T-duality inverts the (dimensionful) coupling of the 2D theory by means of the string slope parameter $\a'$, a fact that from a target space perspective is interpreted as inversion of the radius of the compact space where the degrees of freedom propagate. On the other hand, electric/magnetic duality inverts the (dimensionless) coupling of an Abelian 1-form gauge theory, relating strong and weak coupling regimes. For multiple scalar fields in 2D, the above picture elegantly generalizes to the so-called Buscher rules that relate two geometrically distinct string backgrounds corresponding to the same underlying conformal field theory. Moreover, apart from the above cases, duality is featured in both higher gauge theories, comprising differential forms of degree higher than 1, as well as in theories of gravity at the linearized level.

In this paper, we performed a detailed study of the action of dualities on the couplings of theories that involve multiple fields of the same class and generalized topological theta terms. Our study encompasses general classes of free fields, including scalars, $p$-forms, gravitons and mixed symmetry tensors of degree $(p,1)$, with action functionals that contain a kinetic and a topological sector. Employing a suitable parent action in each case, which includes two independent fields, we showed how on-shell duality relations can be obtained from an off-shell action and by integrating out each independent field in the parent action we determined the two dual, classically equivalent, second-order actions for the dual gauge potentials. As an outcome, we found two sets of Buscher rules for the couplings and showed that the first set \eqref{buschergenmetric} (which is identified with the ordinary rules of \cite{Buscher:1987sk}) applies to theories of multiple  scalars, even degree $2k$-forms and degree $(2k,1)$ mixed symmetry tensors, whereas the second set \eqref{buschergentau} applies to odd degree $(2k-1)$-forms and degree $(2k-1,1)$ mixed symmetry tensors (which includes the graviton and $(\text{dim}\,\S-3,1)$ degree exotic graviton as special cases). The two sets differ in that (i) the topological couplings are antisymmetric for the first and symmetric for the second set, and (ii) they are associated to orthogonal and symplectic transformations respectively.
It is worth highlighting that all the above find a unified origin when a graded-geometric formulation is adopted, thus leading to a universal parent action \eqref{universal}. 

A further interesting outcome is that the gravitational duality relations with $\theta$-term \eqref{HullDR} and \eqref{HullDR2}, introduced in \cite{Hull:2001iu} for single and double duality of the graviton respectively, can be achieved as on-shell solution of a proposed off-shell action. This allowed us to identify which is the precise gravitational theta term that must accompany the linearized Einstein-Hilbert action so that these duality relations are obeyed. It turned out that it is a topological term formed as a product of the Newtonian gravitoelectric field and the gravitomagnetic field of gravitoelectromagnetism, whose nonlinear origin can be traced in quadratic torsion invariants of Nieh-Yan type in spacetimes equipped with the Weitzenb\"ock connection \cite{Chatzistavrakidis:2020wum}.     

One should note that as long as the generalized topological theta angles are constant everywhere, the corresponding terms in the action are total derivatives and may be ignored at the classical level. However, this can potentially change in physical situations where theta angles do not take the same value everywhere. A typical situation of this sort is in the physics of topological insulators, where interfaces separate regions of $\theta=0$ and $\theta=\pi$, both values being consistent with CP or time reversal invariance. Duality of topological insulators has been reported theoretically in \cite{Karch:2009sy,Mathai:2015raa}  and experimentally in \cite{tiexp0,tiexp}. Some theoretical possibilities for analog topological magnetoelectric effects in gravity were considered in \cite{Chatzistavrakidis:2020wum}. It would be interesting to study further physical consequences of this setting in the future, also in light of the results of the present paper.

Our analysis in this paper was restricted to spacetime dimensions where the field strength of the field in question and its dual are of the same degree. Thus the dimension was always $2p+2$ for $p$-forms and $(p,1)$ mixed symmetry tensors. This may be generalized to other dimensions in the same spirit as membrane duality for scalar fields in 3D was studied in \cite{Duff:1990hn}. For example, it would be interesting to explore this generalization for gravitons in 5D, where the dual graviton is the Curtright mixed symmetry tensor of degree $(2,1)$ \cite{Boulanger:2003vs}.  This could be studied in two directions, namely by adding suitable generalized theta term to the action or considering free multi-graviton theories. In the former case, it would be interesting to reconsider the results of \cite{Henneaux:2019zod}, where it was shown that out of the three potential dual fields in 5D gravity, that is the $(1,1)$ graviton, the $(2,1)$ dual graviton and the $(2,2)$ double dual graviton, only two are independent, since the graviton and its double dual are algebraically related. In 4D, however, the presence of a theta term modifies the relation among the latter fields, rendering it non-local and not algebraic. It is expected that a similar statement persists in 5D. In relation to that, the analogous statement for exotic duals of differential forms \cite{Chatzistavrakidis:2019bxo}, i.e. that the standard and exotic duals are algebraically related, should be revisited when theta terms are turned on. 

\paragraph{Acknowledgements.}  A. Ch. and G. K. would like to thank Peter Schupp for inspiring discussions and collaboration in closely related work. We would also like to thank Larisa Jonke and Marc Henneaux for useful discussions. A. Ch. would like to thank the Erwin Schr\"odinger International Institute for Mathematics and Physics for hospitality and financial support during the program ``Higher Structures and Field Theory''. This work is supported by the Croatian Science Foundation Project ``New Geometries for Gravity and Spacetime'' (IP-2018-01-7615).

\appendix 

\section{Identities and computational details for Section \ref{sec4}}
\label{appa}
In this Appendix we collect a number of identities used for the calculations of Section \ref{sec4} and present the computational details regarding the dualization of the linearized graviton in the presence of theta term presented in Section \ref{sec42}. The same identities and computational steps are also used to derive the results of Section \ref{sec51}. 

Recall that in Section \ref{sec4}, we worked on a graded supermanifold equipped with local coordinates $x^{\mu}$ of degree 0 and $\theta^{\mu},\widetilde{\theta}^{\mu}$ of degree 1. A function on this graded manifold is assigned a degree $(p,q)$, each of the two partial degrees $p$ and $q$ being the eigenvalues of the number operators $\hat{p}=\theta^{\mu}\partial/\partial\theta^{\mu}$ and $\hat{q}=\widetilde{\theta}^{\mu}\partial/\partial\widetilde{\theta}^{\mu}$ respectively. Then the six maps $(\hat{p},\hat{q},\eta,\text{tr},\s,\widetilde{\s})$ that correspond to the number operators, the Minkowski metric, the trace and the two cotraces, form a closed algebra with commutation relations in $d$ dimensions, 
\begin{subequations}\bea
&&[\s,\widetilde\s]=\hat p-\hat q\,,\label{ids1a} \\
 &&{[}\text{tr},\h]=d-\hat p-\hat q\,, \label{ids1b}\\ 
&& {[}\hat p,\s]=\s=-[\hat q,\s]\,, \\ 
&& {[}\hat p,\widetilde\s]=\widetilde\s=-[\hat q,\widetilde\s]\,,\\
 && {[}\hat p,\text{tr}]=-\text{tr}=[\hat q,\text{tr}]\,,\\
 && {[}\hat p, \eta]=\eta=[\hat q,\eta]\,,\eea  \end{subequations}
 with every other commutator vanishing. Moreover, they satisfy the following anticommutation relations with the vector fields $\dd$, $\widetilde\dd$ and $\dd^{\dagger}$, $\widetilde{\dd}^{\dagger}$,  
\begin{subequations}
	\bea
	&& \{\text{tr},\dd\}=\widetilde\dd^\dagger=\{\s,\dd^\dagger\}\,,\label{ids2a}\\
&&\{\text{tr},\dd^\dagger\}=0=\{\s,\widetilde\dd^\dagger\}\,,\label{ids2b}\\
&&\{\h,\dd\}=0=\{\s,\dd\}\,,\label{ids2c}\\ 
&&\{\h,\dd^\dagger\}=\widetilde\dd=-\{\widetilde\s,\dd\} \label{ids2d}\eea
\end{subequations}
as well as the corresponding relations obtained by transposing (exchanging $\theta$s and $\widetilde{\theta}$s in) both sides of the above, e.g.  $\{\text{tr},\dd\}=\widetilde\dd^\dagger$ implies that $\{\text{tr},\widetilde\dd\}=\dd^\dagger$ and so on. 

\paragraph{Proof of Eq. \eqref{onshellDR}.} 

In the main text, we stated that the  field equation \eqref{DR} becomes \eqref{onshellDR} on shell, which further implies the duality relation \eqref{HullDR+}, as already shown there. This can be proven as follows. On shell, $f$ is given as $f=\dd e=\dd (h+\widetilde\s b)$. Using that $\mc O$ is given in the present case by \eqref{Ograviton}, we find that  
\bea 
\mc O f|_{f\,\,\text{on-shell}}&=&(1-\frac 12 \widetilde\s \s)\dd(h+\widetilde\s b)=\dd h+\dd\widetilde\s b-\frac 12 \widetilde\s\s\dd h-\frac 12 \widetilde\s\s\dd\widetilde\s b =\nn\\ 
&\overset{\eqref{ids2c},\eqref{ids2d}}=& \dd h +(-\widetilde\s\dd b-\widetilde\dd b)+\frac 12 \widetilde\s\dd\s h+\frac 12 \widetilde\s\dd\s\widetilde\s b = \nn\\ 
&\overset{\eqref{ids1a}}=&\dd h-\widetilde\s\dd b-\widetilde\dd b+\frac 12 \widetilde\s\dd  (\widetilde\s\s b+2b) =  \nn\\
&=&\dd h-\widetilde\dd b\,, \label{lhsDR}
\eea
where in the third and fourth lines we also used the fact that $h$ and $b$ are irreducible, therefore $\s h=0=\s b$.  On the other hand, using the decomposition \eqref{decomposition}, we compute 
\bea 
-\frac{4}{M_{\text{P}}^2}\left(1-\frac 12 \eta\,\text{tr}\right)\dd^{\dagger}\l&=& \left(1-\frac 12 \eta\,\text{tr}\right)\dd^{\dagger}\left(\ast\widehat h+\eta\ast\widehat b\right)= \nn\\
&=&\dd^{\dagger}\ast\widehat h+\dd^{\dagger}\eta\ast\widehat b-\frac 12 \eta\,\text{tr}\dd^{\dagger}\ast\widehat h-\frac 12 \eta\,\text{tr}\dd^{\dagger}\eta\ast\widehat b= \nn\\
&\overset{\eqref{ids2b},\eqref{ids2d}}=&-\ast\dd \widehat h -\eta\,\dd^{\dagger}\ast\widehat b+\widetilde\dd\ast\widehat b+\frac 12 \eta\,\dd^{\dagger}\,\text{tr}\ast\hat h-\frac 12 \eta\,\text{tr}\dd^{\dagger}\eta\ast\widehat b=\nn\\
&\overset{\eqref{ids2d}}=&-\ast\dd\widehat h +\ast \,\widetilde\dd\,\widehat b-\eta\,\dd^{\dagger}\ast\widehat b+\frac 12 \eta\,\text{tr}\,\eta\,\dd^{\dagger}\ast\widehat b-\frac 12 \eta\,\text{tr}\widetilde\dd\ast\widehat b=\nn\\
&=& -\ast\dd\widehat h+\ast\,\widetilde\dd\,\widehat b\,, \label{rhsDR}
\eea
where in the third line we also used the definition of $\dd^{\dagger}$, in the fourth line the fact that $\ast$ and $\widetilde\dd$ commute and that $\widehat h$ is irreducible, therefore $\dd^{\dagger}\text{tr}\ast\widehat h\propto \ast\dd \s\widehat h=0$, and in the fifth line that 
\bea 
\frac 12 \eta\,\text{tr}\,\eta\,\dd^{\dagger}\ast\widehat b-\frac 12 \eta\,\text{tr}\widetilde\dd\ast\widehat b&\overset{\eqref{ids1b}}=& \frac 12 \eta^2\,\text{tr}\dd^{\dagger}\ast\widehat b+\frac 32\eta\,\dd^{\dagger}\ast\widehat b-\frac 12 \eta\,\text{tr}\widetilde\dd\ast\widehat b = \nn\\ &\overset{\eqref{ids2a}}=& \frac 12 \eta^2\,\text{tr}\dd^{\dagger}\ast\widehat b+\frac 32 \eta\, \dd^{\dagger}\ast\widehat b-\frac 12 \eta\,(-\widetilde\dd\,\text{tr}+\dd^{\dagger})\ast\widehat b= \nn\\
&=&\eta\,\dd^{\dagger}\ast\widehat b\,,
\eea
since the terms involving trace acting on a differential form are identically zero. Then, \eqref{onshellDR} follows directly from \eqref{DR}, taking into account \eqref{lhsDR} and \eqref{rhsDR}, and the proof is compete.

\paragraph{Proof of Eq. \eqref{solution DR}.} 
Here we show that the field equation of $f$, which we brought in the form \eqref{DR} in the main text, is solved for $f$ by \eqref{solution DR}. First, acting on both sides of \eqref{DR} with $\ast$, we obtain
\be \ast\,\mathcal{O}f=-\frac{4g^2}{M_{\text{P}}^2}\ast\left(1-\frac{1}{2}\eta\,\text{tr}\right)\dd^\dagger\lambda+\theta g^2\mathcal{O}f\,.
\ee
Substituting this back into \eqref{DR} gives:
\begin{equation}\label{Of}
\mathcal{O}f=-\frac{4}{M_{\text{P}}^2}\,\frac{g^2}{1+\theta^2g^4}\left(1-\theta g^2\ast\right)\left(1-\frac{1}{2}\eta\,\text{tr}\right)\dd^\dagger\lambda\,.
\end{equation}
One can now check that the operator ${\cal O}$ is invertible and its inverse is  $\mathcal{O}^{-1}=1-\widetilde\sigma \sigma$, as shown in \cite{Chatzistavrakidis:2019len}.   Acting with it on both sides of \eqref{Of} gives:
\begin{equation}\label{f}
f=-\frac{4}{M_{\text{P}}^2}\,\frac{g^2}{1+\theta^2g^4}\left(1-\widetilde\sigma \sigma\right)\left(1-\theta g^2\ast\right)\left(1-\frac{1}{2}\eta\,\text{tr}\right)\dd^\dagger\lambda\,.
\end{equation}
Eq. \eqref{solution DR} directly follows from \eqref{f} with the use of \eqref{rhsDR} and this completes the proof.

\paragraph{Derivation of the Lagrangian \eqref{L_II}.} 
Here we fill in some details on the derivation of the dual Lagrangian \eqref{L_II}. 
First, substituting the duality relation \eqref{duality relation} and its solution \eqref{solution DR} back into \eqref{parent2} gives
\bea
	\widetilde{\mathcal{L}}&=&-\frac{M_{\text{P}}^2}{4}\frac{g^2}{1+\th^2 g^4}\int_{\theta,\widetilde\th}\left[(1-\widetilde\sigma\sigma)(1-\theta g^2\ast)\ast (\dd\widehat h-\widetilde{\dd}\,\widehat{b})\right]\ast\widetilde\ast\left[\dd^{\dagger}\ast\widehat{h}+ \dd^\dagger \eta \ast\widehat{b}\right]\nonumber\\
	&=&-\frac{M_{\text{P}}^2}{4}\frac{g^2}{1+\th^2 g^4}\int_{\theta,\widetilde\th}\left[(1-\widetilde\sigma\sigma)(1-\theta g^2\ast)\ast (\dd\widehat h-\widetilde{\dd}\,\widehat{b})\right]\widetilde\ast\left[\dd\widehat{h}+ \ast\,\dd^\dagger \eta \ast\widehat{b}\right]\,.
\eea
By definition of $\dd^{\dagger}$ and the fact that 
\be\label{seta} \widetilde{\sigma}=(-1)^{p(D-p)}\ast\eta\,\ast \ee
 when acting on an arbitrary $p$-form, which follows from Eqs. \eqref{ids1a} and \eqref{ids1b}, it is easy to show that $\ast\,\dd^\dagger\eta\ast\widehat{b}=\dd\,\widetilde{\sigma}\,\widehat{b}$, thus
\bea
	\widetilde{\mathcal{L}}&=&-\frac{M_{\text{P}}^2}{4}\frac{g^2}{1+\th^2 g^4}\int_{\theta,\widetilde\th}\left[(1-\widetilde\sigma\sigma)(1-\theta g^2\ast)\ast (\dd\widehat h-\widetilde{\dd}\,\widehat{b})\right]\widetilde\ast\left[\dd\widehat{h}+ \dd\,\widetilde{\sigma}\,\widehat{b}\right]\nonumber\\
	&=&-\frac{M_{\text{P}}^2}{4}\frac{g^2}{1+\th^2 g^4}\int_{\theta,\widetilde\th}(1-\widetilde\sigma\sigma)\ast\left[(1-\theta g^2\ast) (\dd\widehat h-\widetilde{\dd}\,\widehat{b})\right]\widetilde\ast\left[\dd\widehat{h}+ \dd\,\widetilde{\sigma}\,\widehat{b}\right]\,.\label{L_II-}
\eea
The dual Lagrangian \eqref{L_II} then follows directly from \eqref{L_II-} and the following lemma. For arbitrary tensors $\o$, $\o'$ of degree $(p,q)$ and $\text{min}\{p,q\}\leq 1$, 
\be \label{integral identity}
\int_{\theta,\widetilde\th}(1-\widetilde{\s}\s)\ast \omega\,\widetilde{\ast}\,\omega'=(-1)^{\k}\int_{\theta,\widetilde\th}\omega\star \omega'\,,\quad \k=q(d-q)+pq+1\,.
\ee
To complete the derivation, we prove this lemma as follows. 
First, as proven in \cite{Chatzistavrakidis:2019len}, for tensors $\omega$ and $\xi$ of degree $(p,q)$ and $(d-p,d-q)$ respectively, the following hold:
\bea \label{integral stars}
\int_{\theta,\widetilde{\theta}} \omega\, \xi=-\int_{\theta,\widetilde{\theta}} \ast\, \omega\ast \xi=-\int_{\theta,\widetilde{\theta}} \widetilde{\ast} \,\omega\,\widetilde{\ast} \,\xi\,,\\ \label{integral trace}
\int_{\theta,\widetilde{\theta}} \eta\, \omega\ast\widetilde\ast \,\omega'=\int_{\theta,\widetilde{\theta}} \omega\ast\widetilde{\ast}\,\text{tr} \, \omega'\,.
\eea
In addition, using the definition of $\s$ and Eq. \eqref{seta}, we can easily see that the following relation holds true:
\be \label{sigma identity}
\ast\,\widetilde\s \s\,\ast=-(-1)^{p(d-p)}\h\,\text{tr}\,.
\ee
Then we compute
\bea
	\int_{\theta,\widetilde\th}(1-\widetilde{\s}\s)\ast \omega\,\widetilde{\ast}\,\omega'
	&\overset{\eqref{integral stars}}{=}&-\int_{\theta,\widetilde{\theta}}\ast^2\omega\ast\widetilde{\ast}\,\omega'+\int_{\theta,\widetilde{\theta}}\ast\,\widetilde{\s}\s\ast \omega\,\ast\widetilde{\ast}\,\omega'\nonumber\\
	&\overset{\eqref{sigma identity}}{=}&(-1)^{p(d-p)}\int_{\theta,\widetilde{\theta}}\omega\ast\widetilde{\ast}\,\omega'-(-1)^{p(d-p)}\int_{\theta,\widetilde{\theta}}\h\,\text{tr}\omega\,\ast\widetilde{\ast}\,\omega'\nonumber\\
	&\overset{\eqref{integral stars},\eqref{integral trace}}{=}&(-1)^{p(d-p)}\int_{\theta,\widetilde\theta}\omega\ast\widetilde\ast \,\left(1-\h\,\text{tr}\right)\omega'\,,
\eea
and \eqref{integral identity} follows from the relation between $\star$ and $\ast\widetilde\ast$ in Eq. \eqref{starrelations}.

\end{document}